\documentclass[structabstract,a4paper]{aa}

\usepackage{graphicx}
\usepackage{txfonts}
\usepackage{natbib}
\usepackage{color}

\newcommand{\source}{FIR~4}

\newcommand{\clratio}{[$^{35}$Cl]/[$^{37}$Cl]}

\newcommand{\hhcl}{H$^{37}$Cl}
\newcommand{\htwoclplus}{H$_{2}$Cl$^{+}$}
\newcommand{\htwohclplus}{H$_{2}^{37}$Cl$^{+}$}
\newcommand{\hclplus}{HCl$^{+}$}
\newcommand{\ourhclratio}{$3.2\pm0.1$}
\newcommand{\ourhtwoclplusratio}{$4.3\pm0.8$}

\newcommand{\htwo}{H$_{2}$}
\newcommand{\ohtwo}{ortho-H$_{2}$}
\newcommand{\phtwo}{para-H$_{2}$}

\newcommand{\htwoo}{H$_{2}$O}
\newcommand{\cotwo}{CO$_{2}$}

\newcommand{\nhtwo}{$n_{\rm H_{2}}$}

\newcommand{\meth}{CH$_{3}$OH}

\newcommand{\msol}{$M_{\odot}$}

\newcommand{\lsol}{$L_{\odot}$}

\newcommand{\tkin}{$T_{\textrm{kin}}$}
\newcommand{\tgas}{$T_{\textrm{gas}}$}

\newcommand{\ngas}{$n_{\textrm{gas}}$}
\newcommand{\vlsr}{v$_{\textrm{lsr}}$}

\newcommand{\herschel}{\emph{Herschel}}
\newcommand{\herhifi}{\emph{Herschel}/HIFI}
\newcommand{\hso}{\emph{Herschel} Space Observatory}

\def\tkin{$T_{\rm kin}$}

\begin{document}

   \title{Depletion of chlorine into HCl ice in a protostellar core}

   \subtitle{The CHESS spectral survey of OMC-2 FIR 4}

   \author{
   M.~Kama\inst{1}
	\and
	E.~Caux\inst{2,3}
	\and
	A.~L\'{o}pez-Sepulcre\inst{4,5}
	\and
	V.~Wakelam\inst{6,7}
	\and
	C.~Dominik\inst{8,9}
	\and
	C.~Ceccarelli\inst{4,5}
	\and
	M.~Lanza\inst{10}
	\and
	F.~Lique\inst{10}
	\and
	B.B.~Ochsendorf\inst{1}
	\and
	D.C.~Lis\inst{11,12,13}
	\and
	R.N.~Caballero\inst{14}
	\and
	A.G.G.M.~Tielens\inst{1}
	}

\institute{Leiden Observatory, P.O. Box 9513, NL-2300 RA, Leiden, The Netherlands, \email{mkama@strw.leidenuniv.nl}
        \and
	Universit\'e de Toulouse, UPS-OMP, IRAP, Toulouse, France
	\and
	CNRS, IRAP, 9 Av. colonel Roche, BP 44346, 31028 Toulouse Cedex 4, France
        \and
	Universit\'{e} de Grenoble Alpes, IPAG, F-38000 Grenoble, France
	\and
	CNRS, IPAG, F-38000 Grenoble, France
	\and
	Univ. Bordeaux, LAB, UMR 5804, F-33270, Floirac, France
	\and
	CNRS, LAB, UMR 5804, F-33270, Floirac, France
	\and
	Astronomical Institute Anton Pannekoek, Science Park 904, NL-1098 XH Amsterdam, The Netherlands
        \and
	Department of Astrophysics/IMAPP, Radboud University Nijmegen, Nijmegen, The Netherlands
	\and
	LOMC - UMR 6294, CNRS-Universit\'{e} du Havre, 25 rue Philippe Lebon, BP 1123 - 76 063 Le Havre cedex, France
	\and
	LERMA, Observatoire de Paris, PSL Research University, CNRS, UMR 8112, F-75014, Paris, France
	\and
	Sorbonne Universit\'{e}s, Universit\'{e} Pierre et Marie Curie, Paris 6, CNRS, Observatoire de Paris, UMR 8112, LERMA, Paris, France
	\and
	California Institute of Technology, Cahill Center for Astronomy and Astrophysics 301-17, Pasadena, CA 91125, USA
	\and
	Max-Planck-Institut f\"{u}r Radioastronomie, Auf dem H\"{u}gel 69, 53121 Bonn, Germany 
}

   \date{}

 
  \abstract
   {The freezeout of gas-phase species onto cold dust grains can drastically alter the chemistry and the heating-cooling balance of protostellar material. In contrast to well-known species such as carbon monoxide (CO), the freezeout of various carriers of elements with abundances $<10^{-5}$ has not yet been well studied.}
   {Our aim here is to study the depletion of chlorine in the protostellar core, OMC-2~FIR~4.}
   {We observed transitions of HCl and H$_{2}$Cl$^{+}$ towards OMC-2~FIR~4 using the \emph{Herschel} Space Observatory and Caltech Submillimeter Observatory facilities. Our analysis makes use of state of the art chlorine gas-grain chemical models and newly calculated HCl-H$_{2}$ hyperfine collisional excitation rate coefficients.}
   {A narrow emission component in the HCl lines traces the extended envelope, and a broad one traces a more compact central region. The gas-phase HCl abundance in FIR~4 is $9\times 10^{-11}$, a factor of only $10^{-3}$ that of volatile elemental chlorine. The H$_{2}$Cl$^{+}$ lines are detected in absorption and trace a tenuous foreground cloud, where we find no depletion of volatile chlorine.}
   {Gas-phase HCl is the tip of the chlorine iceberg in protostellar cores. Using a gas-grain chemical model, we show that the hydrogenation of atomic chlorine on grain surfaces in the dark cloud stage sequesters at least $90$\% of the volatile chlorine into HCl ice, where it remains in the protostellar stage. About $10$\% of chlorine is in gaseous atomic form. Gas-phase HCl is a minor, but diagnostically key reservoir, with an abundance of $\lesssim10^{-10}$ in most of the protostellar core. We find the \clratio\ ratio in OMC-2~FIR~4 to be $3.2\pm0.1$, consistent with the solar system value.}
   \maketitle

\section{Introduction}

The freezeout and desorption of volatiles are amongst the key factors that determine the gas-phase abundances of chemical species in interstellar gas, in particular in protostellar cores \citep[e.g.][]{BerginLanger1997, CaselliCeccarelli2012}. The classical example is CO, which is strongly depleted in prestellar cores, but returns to the gas in the warm protostellar stages Class~0 and I \citep[e.g.][]{Casellietal1999, Bacmannetal2002, Jorgensenetal2005}. Similar behaviour, with quantitative differences, is seen or expected for many other species. Here, we investigate the freezeout and main reservoirs of chlorine in a protostellar core.

An ice mantle on a dust grain in a protostellar core mainly consists of a few dominant ice species (e.g. \htwoo, CO, \cotwo, \meth). However, the ices also contain a large number of minor species, accreted from the gas and formed in the ice matrix. Such minor species very likely include HF and HCl, which are the main molecular carriers of the halogen elements fluorine and chlorine, in well-shielded molecular gas. With increasing temperature, such as seen during infall towards a central protostar, the ice mantle desorbs. For any given ice species, the desorption may have multiple stages, with details influenced by its abundance and interaction with the other species in the ice \citep{Collingsetal2004, Lattelaisetal2011}. In the hot cores of protostars, where ice mantles largely desorb, most volatile species are expected to be back in the gas phase.

\begin{table*}[~ht]
\caption{A summary of the observations. Only lines with a flux signal to noise of $\geq5$ are given as detections. All uncertainties and limits are at $1\ \sigma$ confidence. Upper limits for \htwoclplus\ are calculated over $5$~km/s. $^{\star}$ -- \herschel\ observation identifier.}
\label{tab:observations}      
\centering
\begin{tabular}{c c c c c c c}
\hline
\hline
Transition 			& Frequency	& $E_{u}$	& Telescope	& Beam	& Flux			& Notes \& Obsids$^{\star}$	\\
				& [GHz]		& [K]		&			& [\arcsec]		& [K$\cdot$km/s]	& 	\\
\hline
\multicolumn{7}{c}{HCl}	\\
\hline
$1-0$			& $625.9$		& $30.1$	& HIFI	& $34$	&  $2.79\pm0.03$	&1342191591, 1342239639	 \\
				&$625.9$		& $30.1$	& CSO	& $11$	&  $4\pm2$			& -- \\
$2-1$			&$1251.5$	& $90.1$	& HIFI	& $17$	& $1.46\pm0.12$	& 1342216386, 1342239641 \\
\hline
\multicolumn{7}{c}{\hhcl}	\\
\hline
$1-0$			&$625.0$		& $30.1$	& HIFI	& $34$	& $0.88\pm0.03$	& 1342191591, 1342239639 \\
$2-1$			&$1249.5$	& $90.1$	& HIFI	& $34$	& $\leq2.23$		& 1342216386, 1342239641. Blended with CH$_{3}$OH.\\
\hline
\multicolumn{7}{c}{\htwoclplus}	\\
\hline
$1_{1,1}-0_{0,0}$	&$485.4$		& $23.0$	& HIFI & $44$	& $-0.21\pm0.02$		& 1342218633 \\
$2_{0,2}-1_{1,1}$	&$698.6$		& $57.0$	& HIFI & $30$	& $\leq0.05$			& 1342216389 \\
$2_{1,2}-1_{0,1}$	&$781.6$		& $58.0$	& HIFI & $27$	& $-0.62\pm0.05$		& 1342194681 \\
$2_{2,1}-1_{1,0}$	&$1159.2$		& $85.0$	& HIFI & $18$	& $\leq0.19$		& 1342217735 \\
\hline
\multicolumn{7}{c}{\htwohclplus}	\\
\hline
$1_{1,1}-0_{0,0}$	&$484.2$		& $23.0$	& HIFI & $44$	& $\leq0.08$			& 1342218633 \\
$2_{0,2}-1_{1,1}$	&$698.5$		& $57.0$	& HIFI & $30$	& $\leq0.05$			& 1342216389 \\
$2_{1,2}-1_{0,1}$	&$780.0$		& $58.0$	& HIFI & $27$	& $0.26\pm0.06$		& 1342194681 \\
$2_{2,1}-1_{1,0}$	&$1156.0$	& $85.0$	& HIFI & $18$	& $\leq0.19$			& 1342217735 \\
 \hline
\end{tabular}
\end{table*}

We study the depletion of volatile chlorine towards the OMC-2~FIR~4 protostellar core, using HCl and \htwoclplus\ as proxies. We employed the \hso\footnote{Herschel is an ESA space observatory with science instruments provided by European-led Principal Investigator consortia and with important participation from NASA.}\ and the Caltech Submillimeter Observatory (CSO) telescopes and made use of both an updated gas-grain chemical network for chlorine and new calculations of the HCl-\htwo\ hyperfine collisional excitation rate coefficients. In Section~\ref{sec:chem}, we review the interstellar chemistry of chlorine. The source is described in Section~\ref{sec:source} and the observations in Section~\ref{sec:observations}. The analysis separately covers HCl (Section~\ref{sec:hcl}) and \htwoclplus\ (Section~\ref{sec:htwoclplus}), and the results are discussed in Section~\ref{sec:discussion}. We present our conclusions in Section~\ref{sec:conclusions}.

\section{Interstellar chlorine chemistry}\label{sec:chem}

Chlorine has a simple and relatively well characterized interstellar chemistry \citep[e.g.][]{Juraetal1974, Dalgarnoetal1974, Blakeetal1986, Schilkeetal1995, NeufeldWolfire2009}. In dense molecular gas, the gas-phase formation of HCl begins with the reaction 

\begin{equation}
\rm Cl + H_{3}^{+} \rightarrow H_{2}Cl^{+} + H
\label{eq:clplush3plus}
\end{equation}

Dissociative recombination of \htwoclplus\ with $e^{-}$ then produces HCl. A recent study of D$_{2}$Cl$^{+}$ has confirmed the branching ratio of this recombination to be $\sim10$\% into HCl, $\sim90$\% into Cl \citep{Novotnyetal2012}. Below $\sim100$~K, the resulting fraction of chlorine in HCl is $\sim0.1$ to $0.3$ \citep{Dalgarnoetal1974, Blakeetal1986, Schilkeetal1995, NeufeldWolfire2009}. At temperatures of $\gtrsim350$~K, all chlorine can be converted into HCl in dense ($n\gtrsim10^{4}$~cm$^{-3}$) molecular gas by the reaction

\begin{equation}
\rm Cl + H_{2} \rightarrow HCl + H,
\label{eq:clplush2}
\end{equation}

\noindent
on a timescale of $\lesssim 10^{4}$~yr. At colder temperatures, this reaction is much slower.

In photon-dominated regions (PDRs), a layer can form where the reactions $\rm Cl^{+} + H_{2} \rightarrow HCl^{+} + H$ and $\rm HCl^{+} + H_{2} \rightarrow H_{2}Cl^{+} + H$ lead again via dissociative recombination to HCl. In diffuse PDRs ($n_{\rm H}\sim 10^{3}$~cm$^{-3}$), the HCl and \htwoclplus\ column density ratio is $\sim1$, while for $10^{7}$~cm$^{-3}$ it is $\sim10^{2}$ \citep{NeufeldWolfire2009}. At $A_{\rm V} \lesssim 1$ however, the ratio can be $\sim0.01$, while atomic Cl is the dominant gas-phase reservoir of chlorine. At low densities and $A_{\rm V}\ll 1$, Cl$^{+}$ dominates.

The abundance of Cl with respect to atomic hydrogen in the solar photosphere is [Cl]/[H]~$=3.16\times 10^{-7}$ \citep{Asplundetal2009}. The meteoritic abundance is $1.8\times 10^{-7}$ \citep{Lodders2003}, while in the diffuse interstellar medium (ISM) it is $10^{-7}$, indicating a factor of two depletion into refractory grains or volatile ices \citep{Moomeyetal2012}. For our modelling, we refer to the abundance with respect to molecular hydrogen, X(Cl)~$=$~[Cl]/[H$_{2}$], and adopt a reference value X(Cl)~$=10^{-7}$.

The main stable isotopes of chlorine are $^{35}$Cl and $^{37}$Cl. Their ratio in the Solar System is $3.1$ \citep{Lodders2003}.

\section{The \source\ protostellar core}\label{sec:source}

Our target is a nearby intermediate-mass protostellar core, OMC-2~FIR~4\footnote{Identified on SIMBAD as [MWZ90]~OMC-2~FIR~4.} (hereafter \source), located in Orion, at a distance of d~$\approx420$~pc \citep{Hirotaetal2007,Mentenetal2007}. Its envelope mass is $\sim30$~\msol\ and its luminosity within $\sim 20$'' is $\sim100$~\lsol, while the $\sim1$' scale envelope has an estimated luminosity of $\sim400-1000$~\lsol\ \citep[][the latter is hereafter referred to as C09]{Mezgeretal1990, Furlanetal2014, Crimieretal2009}. Class~0 suggests an age of $\lesssim10^{5}$~yr, and Furlan et al. (2014) propose that FIR~4 is amongst the youngest protostellar cores known. In the far-infrared to millimetre regimes, FIR~4 is undergoing intense study in the \emph{Herschel} key programmes CHESS \citep{Ceccarellietal2010, Kamaetal2010, LopezSepulcreetal2013a, LopezSepulcreetal2013b, Kamaetal2013, Ceccarellietal2014} and HOPS \citep{Adamsetal2012, Manojetal2013, Furlanetal2014}, as well as in a number of ground-based projects.
 
The main physical components of \source\ are the warm, clumpy inner envelope and the cold, extended outer one; a proposed outflow; and a tenuous, heavily irradiated foreground cloud \citep{Crimieretal2009, Kamaetal2013, LopezSepulcreetal2013a, LopezSepulcreetal2013b, Furlanetal2014}. The inner envelope has been resolved into continuum peaks with different luminosities \citep{Shimajirietal2008, Adamsetal2012, LopezSepulcreetal2013b, Furlanetal2014}. These sources appear to share a large envelope. The line profiles and excitation of CO and H$_{2}$O show that FIR~4 harbours a compact outflow \citep{Kamaetal2013, Furlanetal2014, Kamaetalinprep}. There are also two PDRs to consider: the dense outermost envelope \citep[\nhtwo~$= 6\times 10^{5}$~cm$^{-3}$,][]{Crimieretal2009}, and the tenuous foreground cloud \citep[\nhtwo~$\approx 10^{2}$~cm$^{-3}$,][]{LopezSepulcreetal2013a}. Whether or not they are physically connected is unclear. For simplicity, we treat them as separate entities, even though there may be a smooth transition in physical conditions between the two.

\begin{figure*}[!ht]
\includegraphics[clip=,width=1.0\linewidth]{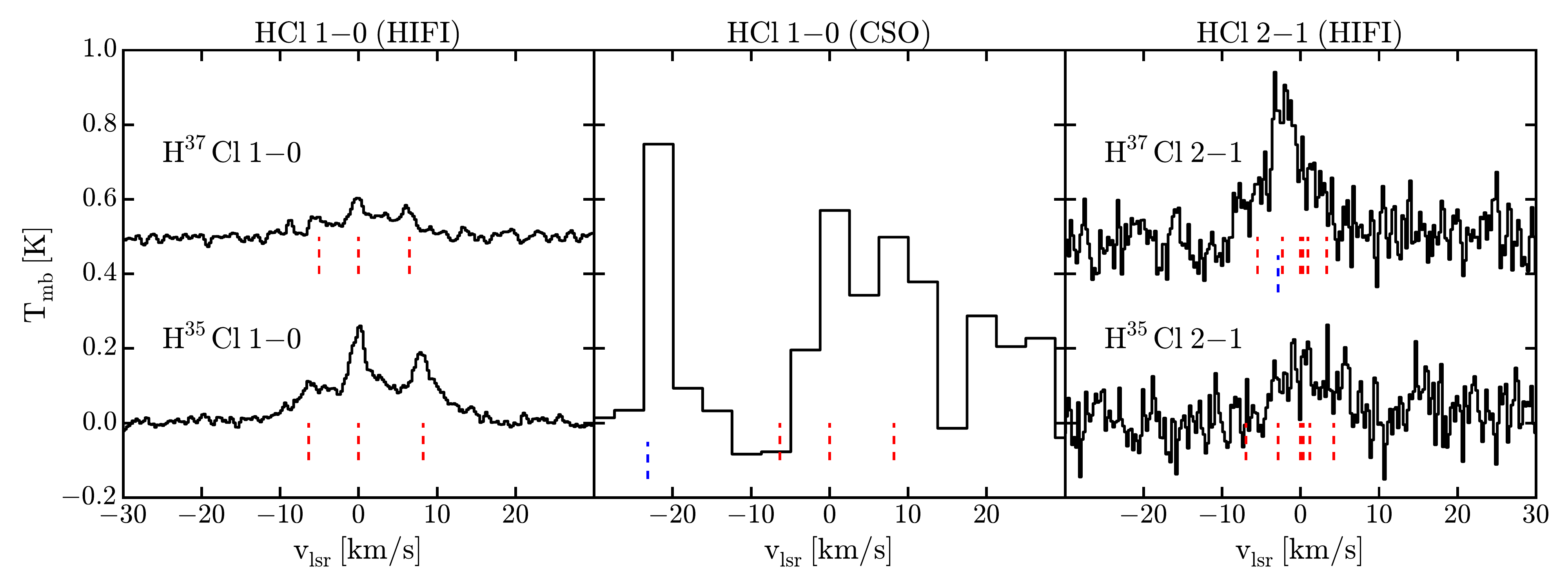}
\caption{The observed spectra of HCl from \herhifi\ and CSO, velocity-corrected and centred on the strongest hyperfine component. Red dashed lines indicate the hyperfine components of the rotational transitions, while blue dashed lines show CH$_{3}$OH lines. The CSO data have been binned to $\delta v=3.7$~km$\cdot$s$^{-1}$.}
\label{fig:observationshcl}
\end{figure*}

\begin{figure}[!ht]
\includegraphics[clip=,width=1.0\linewidth]{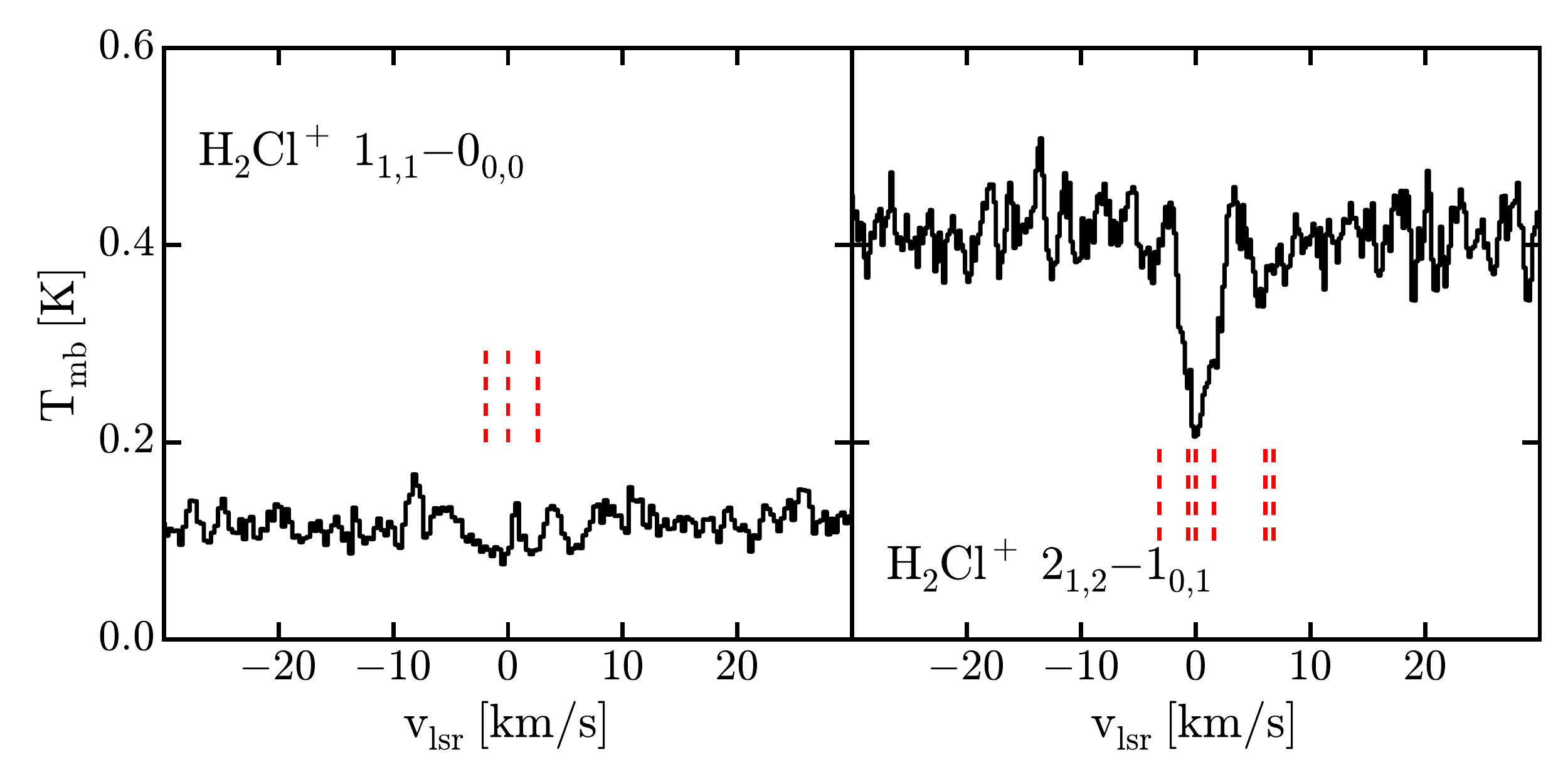}
\caption{The observed spectra of \htwoclplus\ from \herhifi. Only the lines formally detected in absorption are shown, and are velocity-corrected and centred on the strongest hyperfine component. Red dashed lines indicate the hyperfine components of the rotational transitions. The continuum has not been subtracted. See also Fig.~\ref{fig:h2clplus_cassis}.}
\label{fig:observationsh2clplus}
\end{figure}

\section{Observations}\label{sec:observations}

The main neutral and ionized molecular carriers of chlorine in the ISM -- such as HCl, \hclplus\ and \htwoclplus\ -- have strong radiative transitions, although their observations from the ground are hampered by atmospheric water absorption. The observations of HCl and \htwoclplus\ towards \source, summarized in Table~\ref{tab:observations} and shown in Figures~\ref{fig:observationshcl}~(HCl) and~\ref{fig:observationsh2clplus}~(\htwoclplus), were carried out with the \hso\ and the Caltech Submillimeter Observatory (CSO).

\subsection{\herhifi}

As part of the CHESS key programme \citep{Ceccarellietal2010}, \source\ was observed with the Heterodyne Instrument for the Far-Infrared (HIFI) wide band spectrometer \citep{deGraauwetal2010} on the \hso\ \citep{Pilbrattetal2010}, in Dual Beam Switch (DBS) mode at a resolution of $d\nu=1.1$~MHz ($R\sim10^{6}$). Multiple transitions of HCl and \htwoclplus\ were covered. The data quality and reduction, carried out with the HIPE~8.0.1 software \citep{Ott2010}, are presented in \citet{Kamaetal2013}.

\subsection{Caltech Submillimeter Observatory}

Ground-based observations of HCl~$J=1-0$ were carried out in DBS mode with a chopper throw of $240$\arcsec, using the $690$~GHz facility heterodyne receiver of the Caltech Submillimeter Observatory (CSO), on Mauna Kea, Hawaii, on February 6th and 11th, 2013. The atmosphere was characterized by a $225$~GHz zenith opacity of $0.04$-$0.06$ or $1$~mm of precipitable water vapour. Typical single sideband system temperatures were $5000$-$7000$~K. The backend was the high-resolution FFT spectrometer, with $4095$ channels over $1$~GHz of IF bandwidth. The on-source integration time was $51$~min, resulting in an RMS noise of $\sim0.15$~K at a resolution of $3.7$~km/s. Jupiter was used for pointing and calibration. The beam efficiency was 40\%, assuming a $148.5$~K brightness temperature for Jupiter.

\subsection{Overview of the data}

Both HCl and \htwoclplus, as well as their isotopologs, are detected with \herhifi. The HCl~$1-0$ transition is also detected with the CSO. The measured line fluxes, integrated over the hyperfine components, are summarized in Table~\ref{tab:observations}.

The HCl lines show evidence for a broad and a narrow component. Both components are present in the HIFI HCl~$1-0$ data, which also shows a roughly optically thin hyperfine component ratio (2:3:1, in order of increasing $\nu$) for the narrow component. The signal to noise of the HCl~$2-1$ and the CSO $1-0$ data are insufficient to make firm conclusions about the relative importance of the broad and narrow components, however the broad component seems to contribute substantially to both observations. Unfortunately, the H$^{37}$Cl~$2-1$ transition is contaminated by a CH$_{3}$OH transition and is therefore excluded from our analysis.

\section{Analysis of HCl}\label{sec:hcl}

 Here, we first disentangle the kinematical components of the hydrogen chloride lines. Then, we use radiative transfer and chemical modelling to determine the gas-phase HCl abundance. The outermost envelope of \source\ is strongly externally irradiated, and is considered separately in Section~\ref{sec:outerpdr}.

\begin{figure}[!ht]
\includegraphics[clip=,width=1.0\columnwidth]{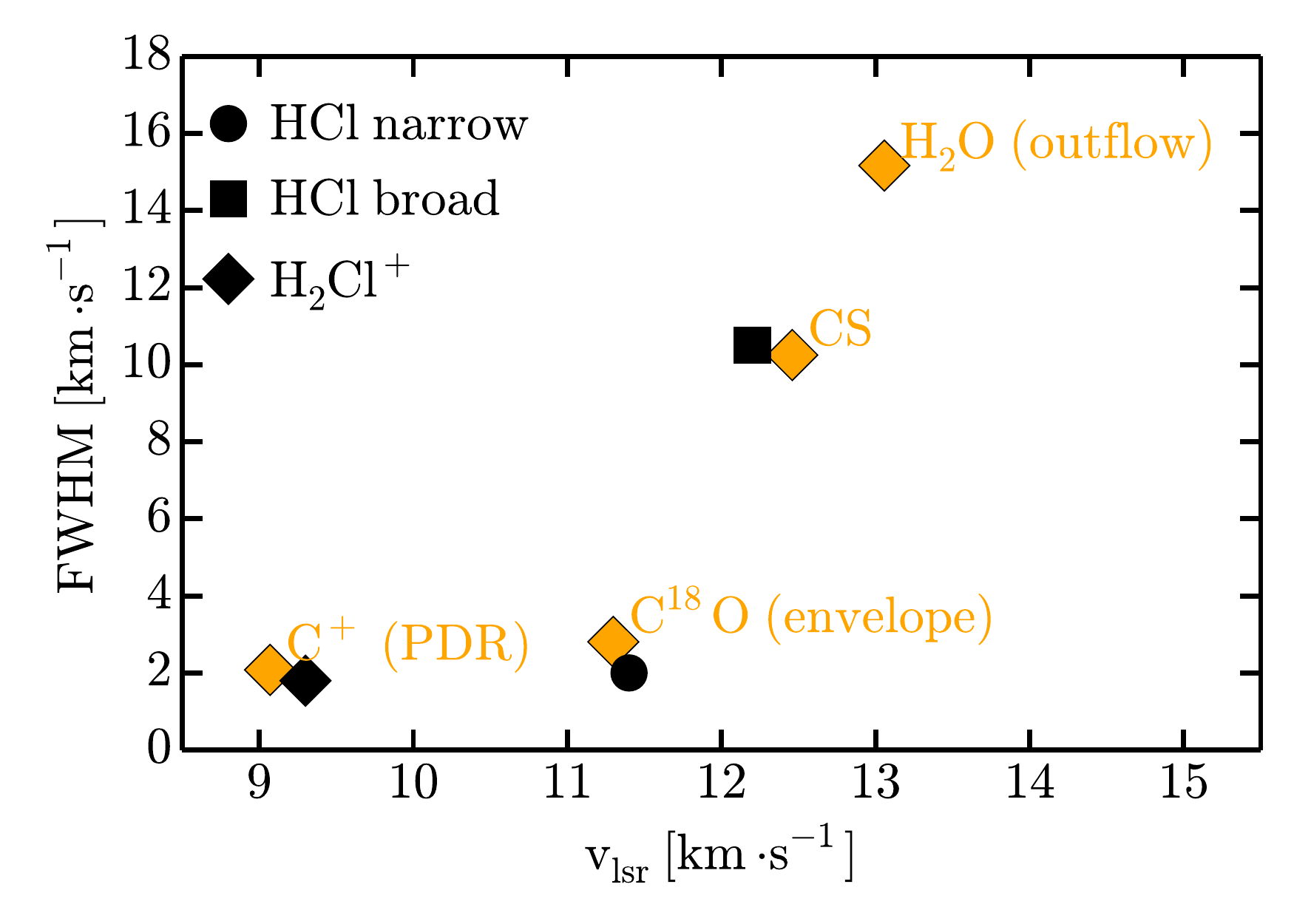}
\caption{The kinematical properties of HCl and \htwoclplus\ compared to other species in \source, each of which is annotated with the main physical component it traces. The label ``PDR'' refers here to the tenuous foreground cloud discovered by \citet{LopezSepulcreetal2013a}. For the comparison species, the mean Gaussian fit parameters from \citet{Kamaetal2013} are plotted.}
\label{fig:kinematics}
\end{figure}

\begin{table}[~ht]
\caption{The kinematical properties of the observed HCl line profiles.}	  
\label{tab:kinematics}
\centering
\begin{tabular}{c c c c}
\hline
\hline
Species	& Component	& \vlsr	& FWHM	\\
		&				& [km$\cdot$s$^{-1}$]	& [km$\cdot$s$^{-1}$] \\
\hline
HCl		& Narrow		& $11.4\pm0.2$	& $2.0\pm0.2	$	\\
HCl		& Broad			& $12.2\pm0.2$	& $10.5\pm0.5$	\\
\end{tabular}
\end{table}

\begin{figure}[!ht]
\includegraphics[clip=,width=1.0\columnwidth]{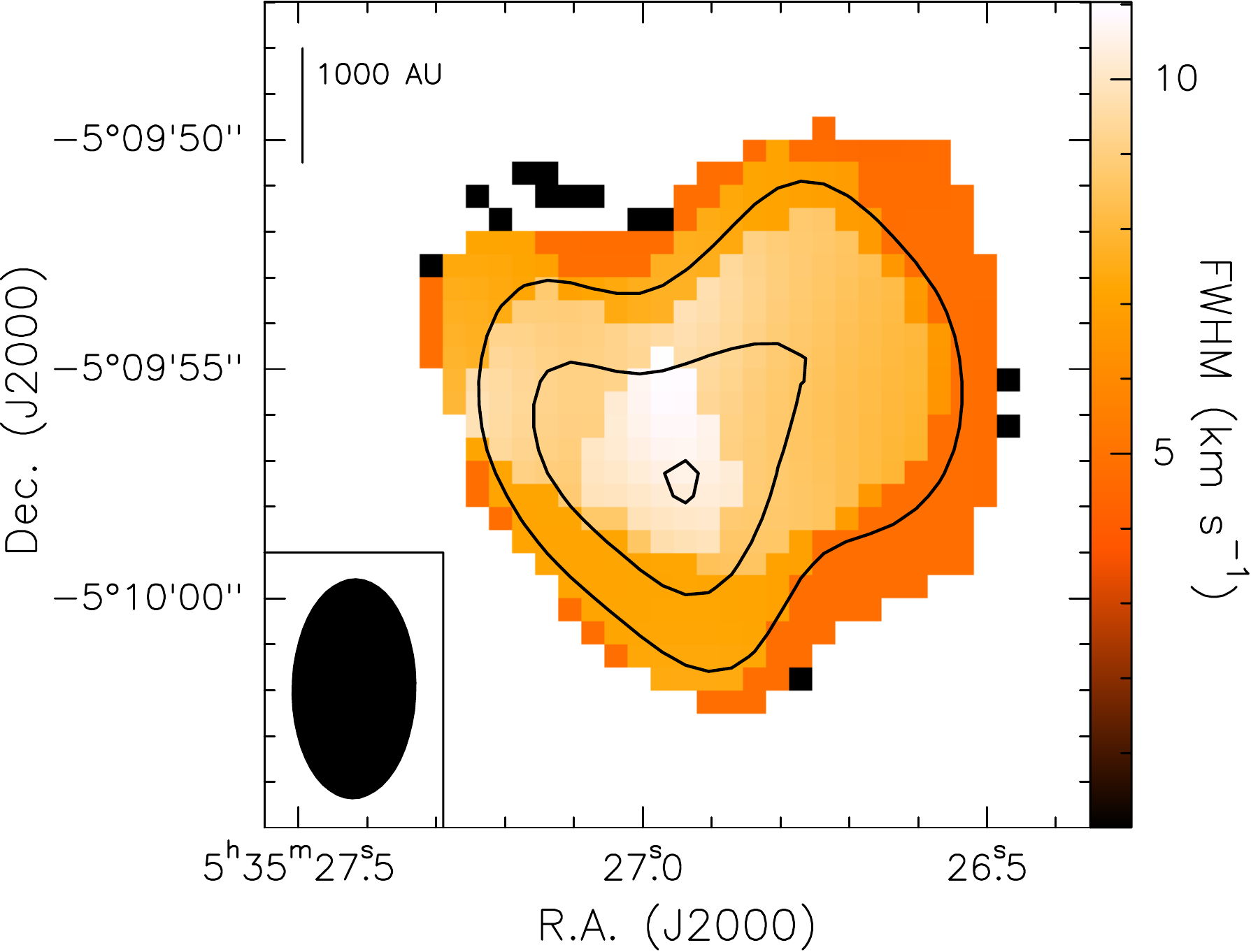}
\caption{A line width map of C$^{34}$S~$3-2$ (colour map), obtained with the Plateau de Bure Interferometer \citep{LopezSepulcreetal2013b}. The channel width of the data is $4$~km$\cdot$s$^{-1}$. Overplotted are continuum contours (black lines), which start at $3\sigma$ with a $3\sigma$ step, with $1\sigma=0.14$~Jy$\cdot$km$\cdot$s$^{-1}\cdot$beam$^{-1}$. North is up and East is left.}
\label{fig:csmap}
\end{figure}

\subsection{Kinematics}\label{sec:hclkinematics}

A two-component model of the hyperfine structure in the HCl~$1-0$ observations strongly constrains the kinematic parameters of the broad and narrow components, given in Table~\ref{tab:kinematics}.

In Figure~\ref{fig:kinematics}, we compare the kinematic components of HCl to those of other species from the \source\ HIFI spectrum of \cite{Kamaetal2013}. The properties of the narrow component match the large-scale quiescent envelope tracers, such as C$^{18}$O. The broad component parameters lie between those of the envelope and the outflow tracers, and match the mean properties of CS.

\begin{figure}[!ht]
\centering
\includegraphics[clip=,width=0.8\columnwidth]{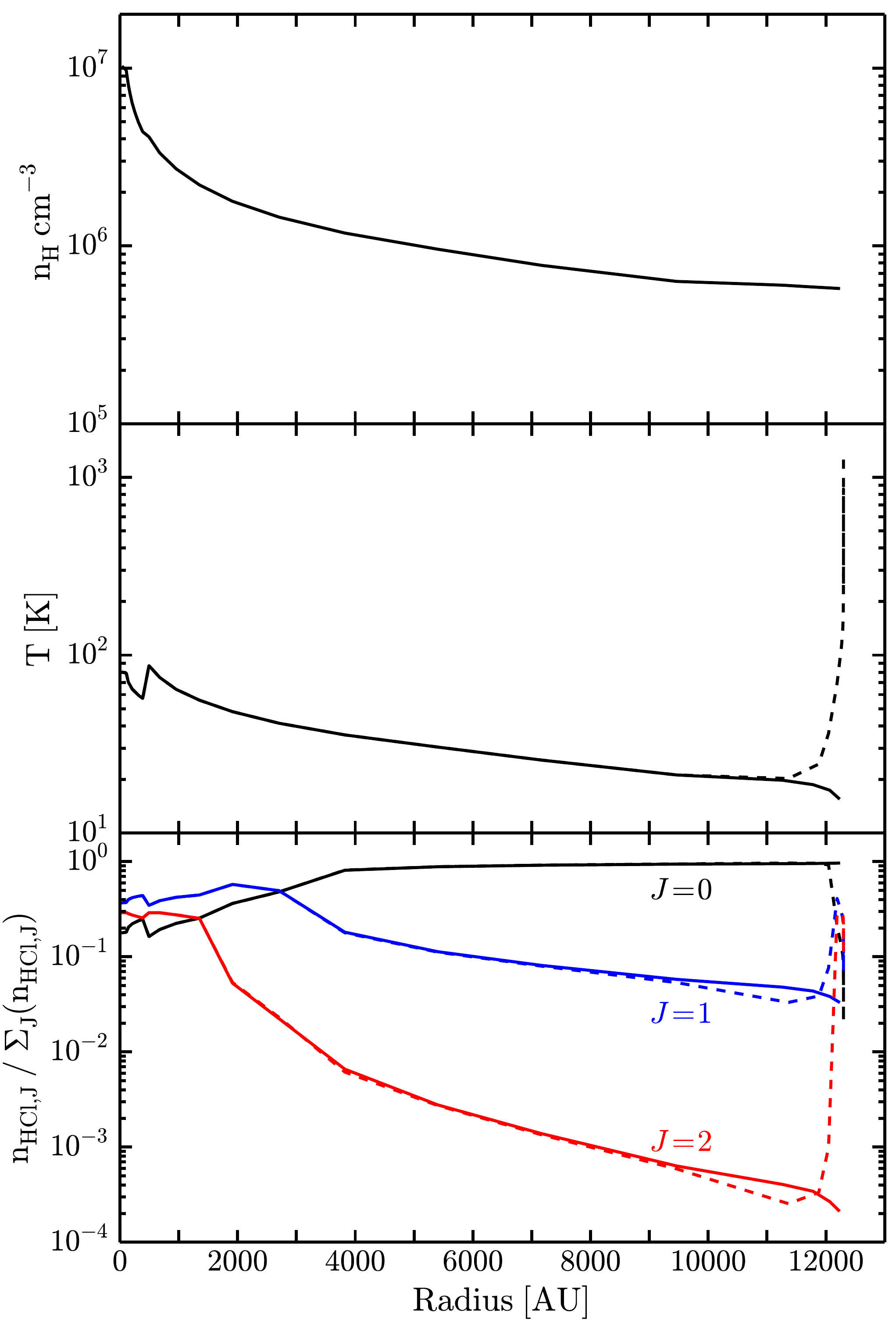}
\caption{\emph{Top and middle panels: }The large-scale source structure of \source\ (\ngas\ at top, \tgas\ at middle) from \citet[][C09]{Crimieretal2009}, corresponding to an integrated luminosity of $1000$~\lsol. \emph{Bottom panel: }  The relative populations of the lowest HCl rotational states, showing $J=0$ (black), $J=1$ (blue), and $J=2$ (red) for the C09 source structure without modification (solid lines) and with an added external irradiation of $G_{0}=415$~ISRF (dashed lines; Section~\ref{sec:outerpdr}). The level populations were modelled with Ratran using the new hyperfine collision rate coefficients presented in Appendix~\ref{apx:collrates}.}
\label{fig:source}
\end{figure}

\begin{figure}[!ht]
\includegraphics[clip=,width=1.0\linewidth]{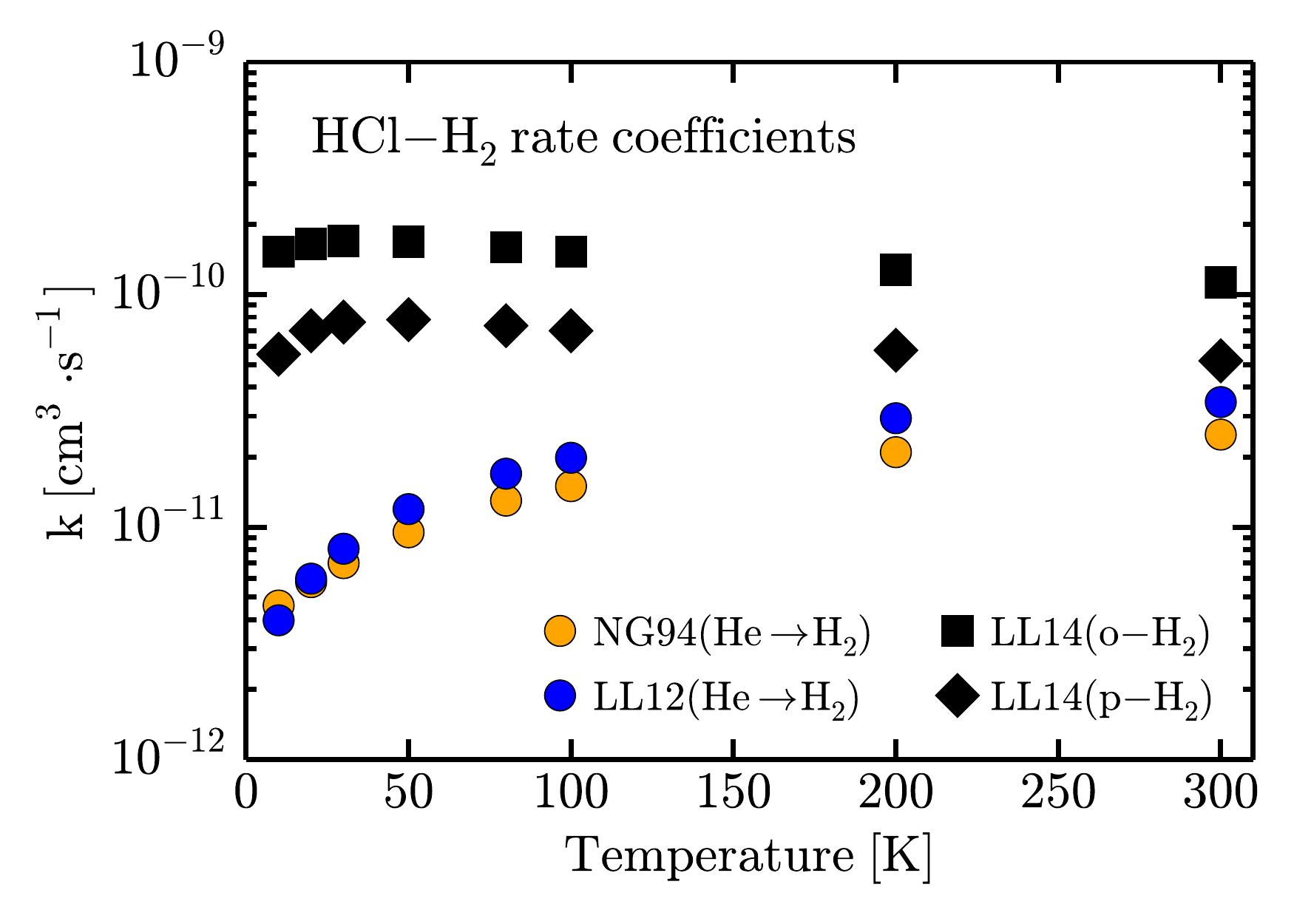}
\caption{The rate coefficients for the collisional excitation HCl $J=0\rightarrow1$ by \htwo. The previously available HCl-He rate coefficients, scaled by $1.38$ to account for the reduced mass difference, are from \citet[][orange circles]{NeufeldGreen1994} and \citet[][blue circles]{lanza_collisional_2012}. The new HCl-\ohtwo\ and -\phtwo\ rate coefficients (Appendix~\ref{apx:collrates}) are shown by black squares and diamonds, respectively.}
\label{fig:collrates}
\end{figure}

Contrary to HCl, for which exceptional observing conditions or space observatories are required, isotopologs of CS are readily observed with ground-based instruments. In Figure~\ref{fig:csmap}, we show a map of the C$^{34}$S~$3-2$ line width, based on Plateau de Bure Interferometer data from \citet{LopezSepulcreetal2013b}. A spatially resolved region $\sim10$\arcsec\ across ($2100$~AU radius) and detected at a $3\sigma$ confidence at $F_{C^{34}S}\geq0.4$~~Jy$\cdot$km$\cdot$s$^{-1}\cdot$beam$^{-1}$ has a line width of $\gtrsim6$~km/s. The spatial resolution of the data is $\sim5$\arcsec, the channel width is $4$~km/s, and the flux loss compared to single-dish data is $\sim60$\%, so we can only conclude that if the C$^{34}$S broad component is equivalent to that of HCl, the latter is centrally peaked and extended on several thousand AU scales.

Based on the arguments presented above, we attribute the narrow and broad HCl line profile component to the outer envelope and some compact yet dynamic inner region, respectively. Given the large difference in line width between the broad HCl component and the outflow tracers CO and \htwoo, it is unlikely that they trace the same volumes of gas.

\begin{figure*}[!ht]
\includegraphics[clip=,width=1.0\linewidth]{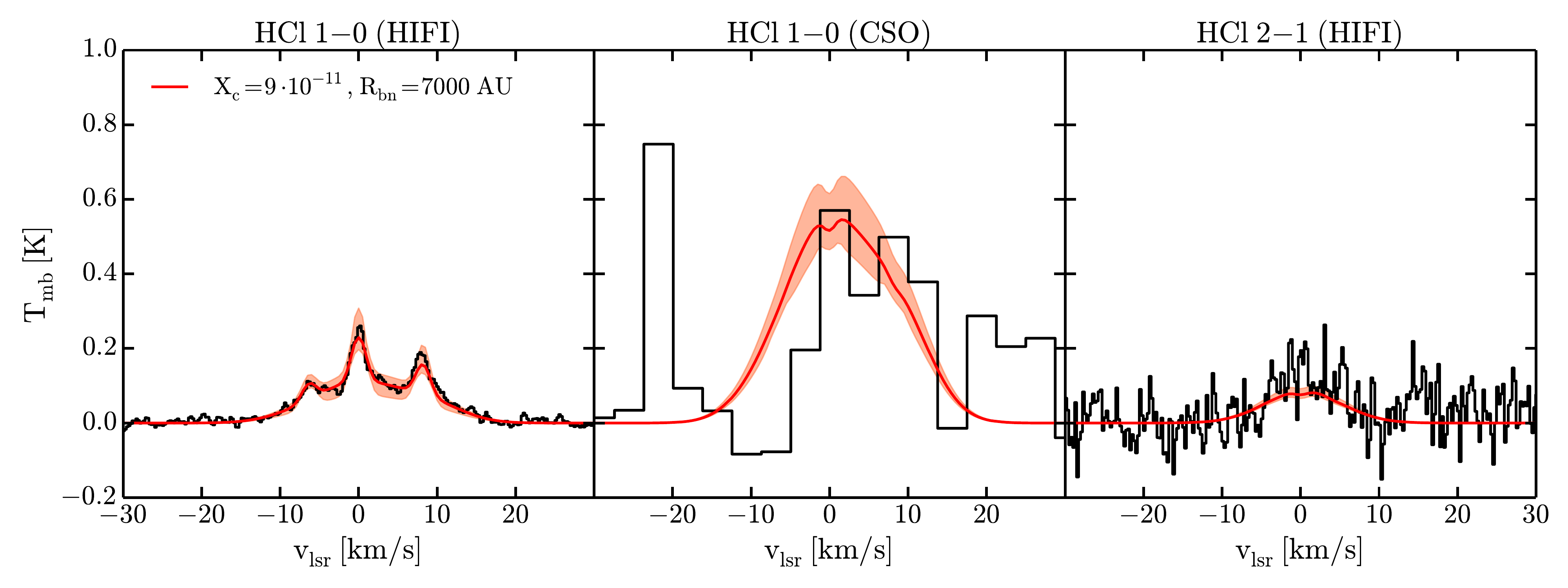}
\caption{The HCl~$1-0$ and $2-1$ transitions as observed with HIFI and CSO (black). Overplotted are the best-fit constant HCl abundance model (red), and the reduced $\chi^{2}<3$ range (shaded orange) of the models. The strongest feature in the CSO data, at $-20$~km$\cdot$s$^{-1}$, is a \meth\ line.}
\label{fig:hclmodels}
\end{figure*}

\begin{figure}[!ht]
\includegraphics[clip=,width=1.0\linewidth]{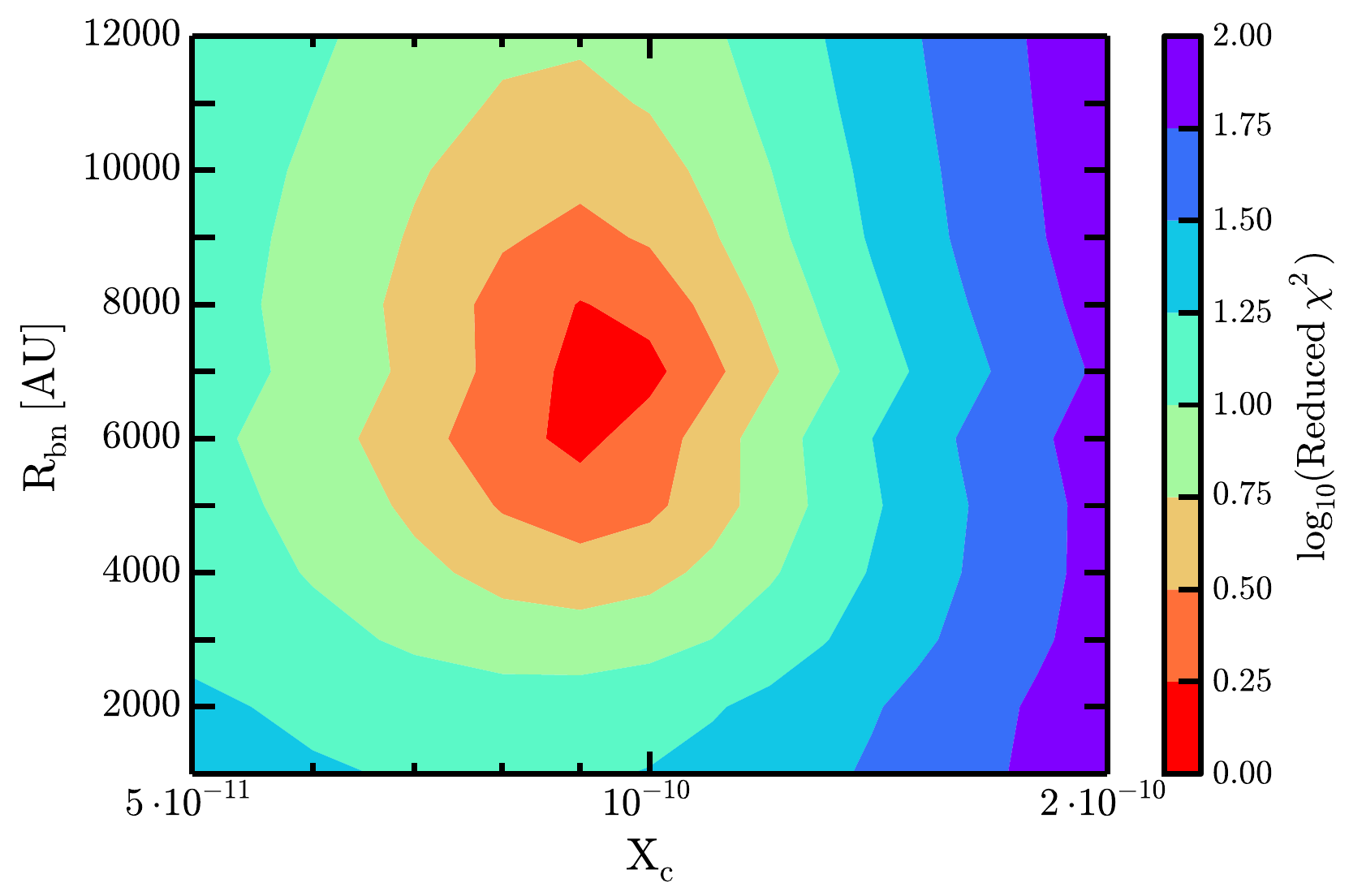}
\caption{The $\rm \log_{10}{(reduced\ \chi^{2})}$ surface for the two constant HCl abundance model parameters: $X_{c}$ and $R_{bn}$. }
\label{fig:chisq}
\end{figure}

\subsection{The gas-phase HCl abundance in \source}\label{sec:crimiermodel}

To obtain an abundance profile, we modelled the HCl excitation and radiative transfer in \source\ using the Monte Carlo code, Ratran\footnote{\texttt{http://www.sron.rug.nl/~vdtak/ratran/}} \citep{HogerheijdevanderTak2000} as described below.

\subsubsection{New HCl-\htwo\ collisional excitation rates}\label{sec:collrates}

We modelled the HCl emission using new hyperfine-resolved collisional excitation rate coefficients, presented in detail in Appendix~\ref{apx:collrates}. These are based upon the recent potential energy surface and rotational excitation rate coefficients of \citet{lanza_near-resonant_2014} and \citet{lanza_taux}. In Figure~\ref{fig:collrates}, we compare the new rate coefficients to the HCl-He ones from \citet{NeufeldGreen1994} and \citet{lanza_collisional_2012}. The latter, scaled to H$_{2}$ collisions by a mass correction factor of $1.38$, differ from the new coefficients by a factor of a few at \tkin~$>50$~K, and by around a factor of ten at \tkin~$<50$~K. A similar difference has been found for HF \citep{GuillonStoecklin2012}. We discuss the impact of the new excitation rates on HCl depletion estimates in Section~\ref{sec:rateimpact}.

\begin{figure*}[!ht]
\includegraphics[clip=,width=0.49\linewidth]{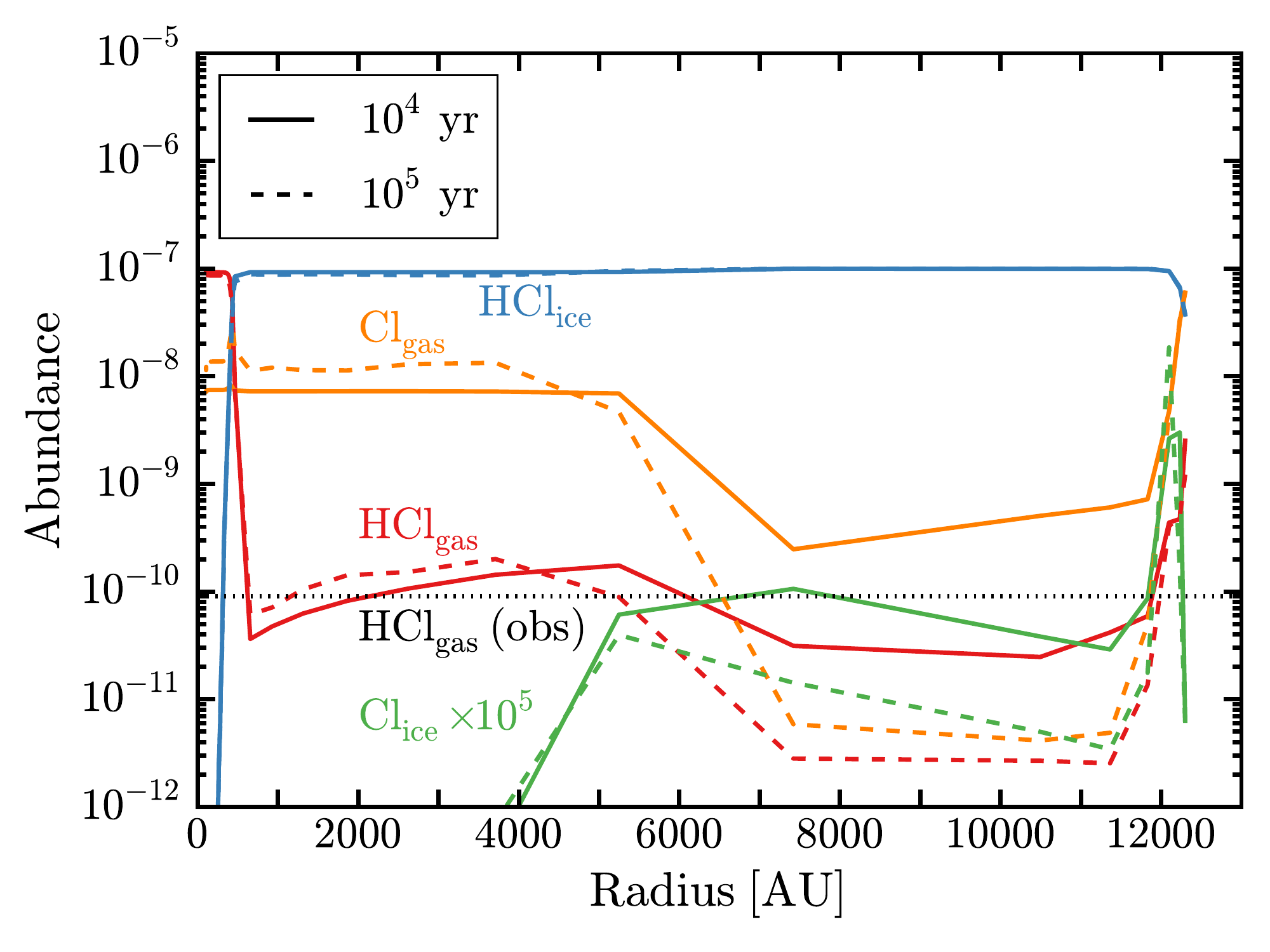}
\includegraphics[clip=,width=0.49\linewidth]{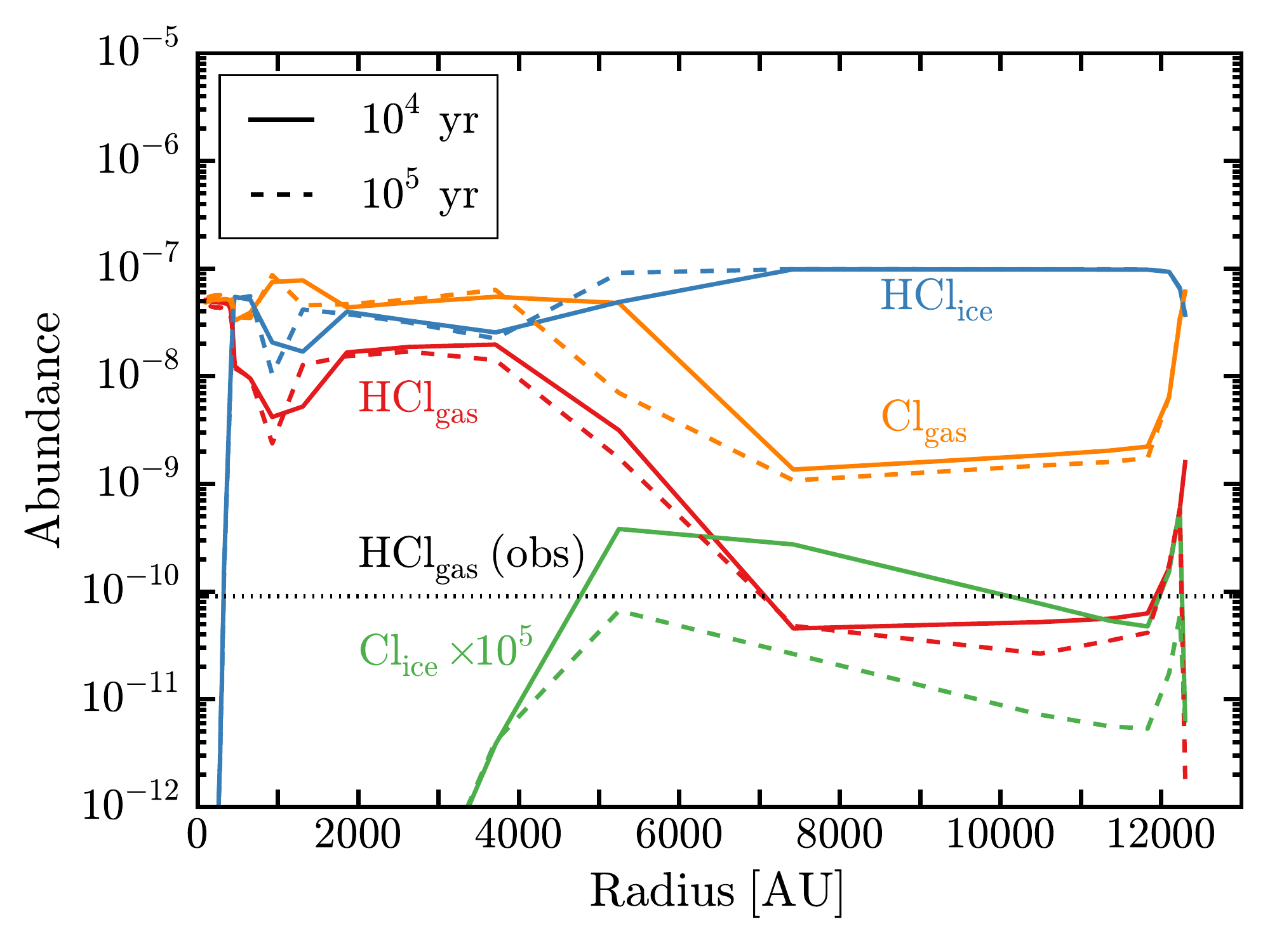}
\caption{The HCl and Cl gas- and ice-phase abundances by $10^{4}$ (solid) and $10^{5}$~yr (dashed) after the dark cloud stage, as modelled with Nautilus. The cosmic ray ionization rate is $\zeta=10^{-16}$~s$^{-1}$ (left panel) and $\zeta=10^{-14}$~s$^{-1}$ (right panel).}
\label{fig:nautilus}
\end{figure*}

\subsubsection{The source model}

As the basis of our modelling, we adopted the spherically symmetric large-scale source structure with no enhancement of the external irradiation field ($G_{0}=1$~ \emph{interstellar radiation fields}, ISRF) from C09. The density and temperature profiles, as well as the relative populations of the relevant HCl rotational levels, are shown in Figure~\ref{fig:source}. Due to the limited spatial resolution of the continuum maps it is based on, the source model is not well constrained on scales $\lesssim2000$~AU. Thus, we interpret it as the spherically-averaged large-scale structure of the source. The total H$_{2}$ column density in a pencil beam through the centre of the source model is $N$(\htwo)~$=4.6\times 10^{23}$~cm$^{-2}$.

According to the relative level populations shown in Fig.~\ref{fig:source}, the $J_{u}=1$ state is most relevant within $\sim4000$~AU, with a factor of five to ten decrease at larger radii, while the $J_{u}=2$ state is mostly populated in the inner $\sim2000$~AU and plays no role in the outer envelope. If an external irradiation field is added (Section~\ref{sec:outerpdr}), the gas temperature reaches $\sim1000$~K in a thin outer layer, and the $J=2$ and higher level population gains in importance. Thus, the HCl~$2-1$ line constrains the abundance within $2000$~AU and more weakly in a thin outer layer, while the $1-0$ transition constrains it in the bulk of the envelope.

\subsubsection{Fitting a constant abundance profile}

As a first guess, we assume a constant HCl abundance in the source. Assuming a fixed source structure, the model has only two free parameters: $X_{c}$, the gas-phase HCl abundance; and $R_{bn}$, the radius where the line width switches from broad to narrow. We performed a $\chi^{2}$ minimization on the HCl~$1-0$ and $2-1$ line profiles from HIFI, and the $1-0$ from CSO. The best fit parameters, with a reduced $\chi^{2}=1.45$, are $X_{c}=9\times 10^{-11}$ and $R_{bn}=7000$~AU. In Figure~\ref{fig:hclmodels}, we show the data, the best-fit model and the range of models within reduced $\chi^{2}=3$. The $\log_{10}{(\chi^{2})}$ surface is shown in Figure~\ref{fig:chisq}.

\subsection{A full chemical model}\label{sec:nautilus}

We modelled the chlorine chemistry at each radial location in \source\ with the Nautilus gas-grain chemical code. Nautilus time-dependently computes the gas and grain chemistry including freezeout, surface chemistry, and desorption due to thermal and indirect processes. The grain surface reactions are described in \citet{Hersantetal2009} and \citet{Semenovetal2010}. The binding energy of Cl on \htwoo\ ice is $1100$~K, for HCl we adopted $5174$~K from \citep{Olanrewajuetal2011}. The gas-phase network is based on \texttt{kida.uva.2011}, from \citet{Wakelametal2012}, and was updated with data from \citet{NeufeldWolfire2009}. The full network contains 8335 reactions, 684 species and 13 elements. The \htwo\ and CO self shielding are computed following \citet{Leeetal1996}, as described in \citet{Wakelametal2012}. We adopted a volatile chlorine abundance of $10^{-7}$. The chemistry is first evolved to steady state in dark cloud conditions. This yields the initial abundances for the time-dependent chemistry in \source, using the C09 source structure for the density and temperature.

\emph{Dark cloud stage. }The initial conditions for the \source\ calculation are computed for dense, cold cloud conditions: a temperature of $10$~K, $n_{\rm H} = 2\times 10^{4}$~cm$^{-3}$, $\rm A_{v}=10$~mag, and a cosmic-ray ionization rate of $\zeta=10^{-17}$~s$^{-1}$. The model is evolved to $10^{6}$~yr. The species are initially all atomic, except for \htwo, with most abundances from \citet{Hincelinetal2011}. The abundance of oxygen is set to $3.3\times 10^{-4}$, and of chlorine to $10^{-7}$. At the end of the dark cloud stage, about $93$\% of elemental chlorine is in HCl ice, the remaining $7$\% is almost entirely in gas-phase atomic Cl. The temperature, the density and the age of the cloud influence the fraction of Cl and HCl in the gas versus in the ices, but do not affect the HCl/Cl ratio. This ratio, both in the gas and ices, is mostly influenced by the cosmic ray ionization rate $\zeta$. The resulting abundances are used as the initial conditions for \source.

\emph{Protostellar stage. }For the protostellar stage, we use the C09 density and temperature structure without enhanced external irradiation. The outermost envelope is treated in more detail in Section~\ref{sec:outerpdr}. In Figure~\ref{fig:nautilus}, we show the gas and ice abundances of HCl and atomic Cl in \source, as modelled with Nautilus at ages $10^{4}$ and $10^{5}$~years. The trial cosmic ray ionization rates were $10^{-16}$~s$^{-1}$ \citep[left panel; a foreground cloud value from][]{LopezSepulcreetal2013a} and $10^{-14}$~s$^{-1}$ \citep[right panel; recently inferred for \source\ by][]{Ceccarellietal2014}.

As seen from the blue lines in Figure~\ref{fig:nautilus}, HCl ice is the main reservoir of elemental chlorine outside of $\sim5000$~AU at both times in both models. It stores $90$ to $98$\% of all elemental chlorine. The second reservoir is gaseous atomic chlorine, at $1$ to $10$\% of the elemental total, typically an order of magnitude above the gas-phase HCl abundance. Inside $5000$~AU, HCl ice contains $90$\% ($\zeta=10^{-16}$~s$^{-1}$) or $20$\% ($\zeta=10^{-14}$~s$^{-1}$) of the chlorine. The HCl ice is built up over the $10^{6}$~year cold dark cloud stage. The gas-phase HCl abundance slowly decreases with time in most of the source, due to continuing freezeout.

Above, the physical structure switched instantly from the dark cloud structure to the protostellar core. Adding time dependency to the density structure would decrease the abundance changes in the inner envelope, which would be replenished with pristine material from the outer envelope. Additionally, using a higher prestellar core density would lead to even stronger and more rapid freezeout of chlorine into HCl ice, increasing the depletion.

The gas-phase abundance of HCl from Nautilus is consistent with the observational constraints on $X$(HCl)$_{gas}$, as reported in Section~\ref{sec:crimiermodel}, within a factor of a few. This is with the exception of the high-ionization model, where $X$(HCl)$_{gas}$ exceeds the observed limit by two orders of magnitude. The gas-phase HCl abundance of $\sim10^{-10}$ -- seen in most of the source in both models -- is the tip of the iceberg, as $90$\% of all chlorine is frozen onto grain surfaces as HCl and likely in a water ice matrix.

\begin{figure}[!ht]
\includegraphics[clip=,width=0.500\columnwidth]{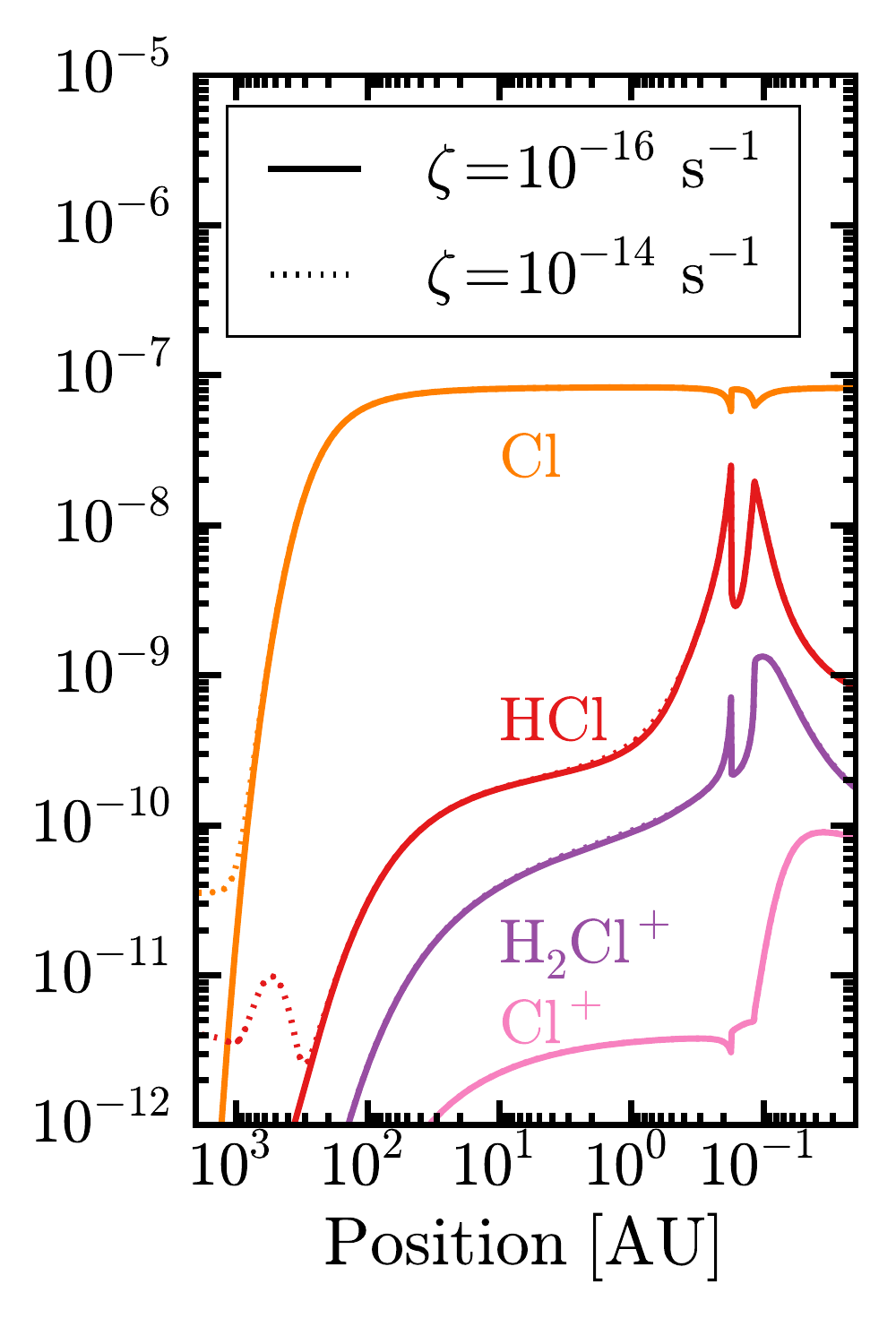}
\includegraphics[clip=,width=0.485\columnwidth]{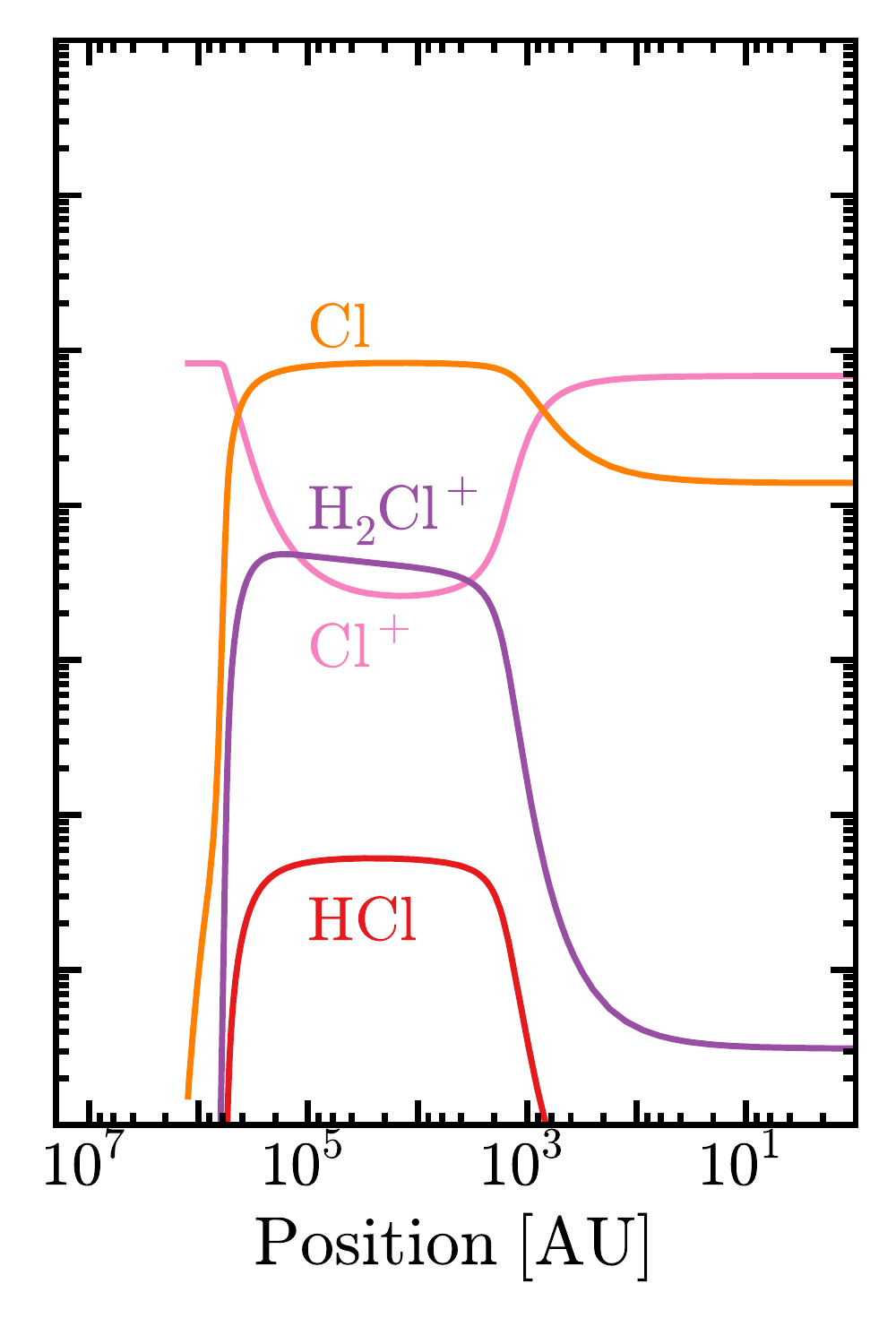}
\caption{The Meudon models of the chlorine chemistry in the two PDR layers seen towards \source, showing abundances relative to \ngas. \emph{Left panel: }The dense PDR in the outermost envelope of \source, with \ngas~$=6\times 10^{5}$~cm$^{-3}$, $G_{0}=415$~ISRF. The cosmic ray ionization rate is $\zeta=10^{-16}$~s$^{-1}$ (solid lines) and $\zeta=10^{-14}$~s$^{-1}$ (dotted). The position is given outside-in from the right. \emph{Right panel: }The tenuous PDR in the foreground of \source, with \ngas~$=10^{2}$~cm$^{-3}$, $G_{0}=1500$~ISRF and $\zeta = 3\times 10^{-16}$~s$^{-1}$. The position is given from the strongly irradiated side on the right.}
\label{fig:meudon}
\end{figure}

\subsection{The outermost envelope PDR}\label{sec:outerpdr}

The dense and heavily irradiated photon dominated region in the outermost envelope requires a specialized treatment. We employ the Meudon\footnote{\texttt{http://pdr.obspm.fr/PDRcode.html}} PDR code \citep{LePetitetal2006} for this. The physical structure is a slab with a constant density of $n_{H}=6\times 10^{5}$ -- the outermost density in the C09 source model -- extending to $A_{\rm V}=20$~mag. We used two cosmic ray ionization rates, as before: $\zeta=10^{-16}$~s$^{-1}$ and $10^{-14}$~s$^{-1}$.

The external irradiation of \source\ has previously been inferred to be $G_{0}\sim500$ interstellar radiation field units, based on [CII]~$158\ \mu$m emission \citep{Herrmannetal1997}. We re-evaluated $G_{0}$ using the \herhifi\ [CII] line flux from \citet{Kamaetal2013}, which was obtained at higher spatial resolution and yields $G_{0}\sim300$. We further checked $G_{0}$ based on an archival \emph{Spitzer} $8 \mu$m map, which has a substantial contribution from PAH emission, and this directly counts excitation events by ultraviolet photons. The measured flux density of $\sim250$~MJy/sr gives $G_{0}=415$, using Eq.~1 of \citet{VicenteBerneTielens2013} with a standard [C]/[H]$\sim10^{-4}$ and $4$\% of elemental carbon locked in PAHs \citep{Tielens2005}. All estimates are consistent with each other, and with irradiation by the Trapezium cluster at a projected separation of $\sim2$~pc. We adopt the middle ground: $G_{0}=415$.

The resulting abundance profiles of gas-phase HCl, Cl, Cl$^{+}$ and \htwoclplus\ are shown in the left-hand panel of Figure~\ref{fig:meudon}. For both $\zeta$ values, HCl peaks at $\sim10^{-8}$ at the H/\htwo\ transition, then decreases deeper into the PDR. By $A_{\rm V}\approx 1$, $X$(HCl)$_{gas}$ drops below $10^{-10}$. The HCl abundance in these models is different from the Nautilus results because of different initial conditions and because the Meudon code solves for steady state, which the time-dependent Nautilus models do not reach by $10^{5}$~yr.  The rapid changes around $0.1$ to $0.2$~AU are due to numerical issues related to \htwo\ formation and the heating-cooling balance. The effect on HCl is limited to $\lesssim40$\% of the column density, and does not substantially affect our conclusions. The temperature in this region is between $\sim100$ and $\sim1000$~K, and the total column density in the outer PDR is N(HCl)~$\approx1\times 10^{12}$~cm$^{-2}$.

Using the $X$(HCl) profile shown in Figure~\ref{fig:meudon} for the outermost envelope, we re-fitted the HCl abundance model from Section~\ref{sec:crimiermodel}. The two $X$(HCl) profiles were joined at \tkin~$=25$~K. The impact of the PDR on the HCl emission is small. The best-fit combined model -- $X_{c}=7\times 10^{-11}$, $R_{bn}=9000$~AU -- has a reduced $\chi^{2}=1.55$. The model matches a tentative narrow peak in the HCl~$2-1$ line, which the constant abundance model does not, although this does not improve the global $\chi^{2}$.

In the outer envelope PDR, elemental chlorine is not depleted from the gas in the outermost $300$~AU. The PDR model is consistent with the observed HCl emission, and still requires most of the narrow and broad HCl flux to originate deeper in \source. It is also consistent with the upper limit on \htwoclplus\ absorption at $11.4$~km/s. The \htwoclplus\ data is analysed next, in Section~\ref{sec:htwoclplus}.

\begin{table}[!ht]
\begin{center}
\caption{The observational results and model predictions for \htwoclplus\ in the tenuous foreground PDR. Dashes are quantities not obtainable from the chemical model. See Section~\ref{sec:htwoclplus} for details.}
\label{tab:htwoclplus}
\begin{tabular}{ c c c c c }
			& $\rm N$~[$10^{14}$~cm$^{-2}$]	& $\rm T_{ex}$~[K]	& \vlsr~[km/s]	& $\delta v$~[km/s]	\\
\hline
\hline
Observed	& $1.3\pm0.1$	& $4.3\pm0.1$	& $9.3\pm0.1$	& $1.8\pm0.1$	\\
Modelled	& $0.11$			& --			& 	--		& --			\\
\hline
\end{tabular}
\end{center}
\end{table}

\section{Analysis of \htwoclplus}\label{sec:htwoclplus}

The \htwoclplus\ lines, shown in Figure~\ref{fig:h2clplus_cassis}, appear in absorption against the weak continuum of the source. As shown in Figure~\ref{fig:kinematics}, they are blueshifted by $2$~km/s with respect to the systemic velocity of \source, which is $11.4$~km/s. Thus, the \htwoclplus\ absorption does not originate in the dense outer envelope PDR. Instead, the radial velocity matches that of the tenuous foreground PDR layer recently identified towards \source\ by \citet{LopezSepulcreetal2013a}. The distance and relation between the foreground layer and FIR~4 are poorly constrained.

We used the CASSIS\footnote{\texttt{http://cassis.irap.omp.eu}} software to perform Markov Chain Monte Carlo fitting of local thermodynamic equilibrium models to the \htwoclplus\ and \htwohclplus\ lines. Not all the lines are detected, but are still useful as constraints. To minimize baseline issues, we added the continuum fit of \citet{Kamaetal2013} to baseline subtracted spectra. We find a column density of $N$(\htwoclplus)$~=~1.3\times 10^{14}$ and an isotopic ratio of \clratio~$=(4.3\pm0.8)$. The main fitting results are given in Table~\ref{tab:htwoclplus}.

To compare the observed \htwoclplus\ column density with that expected from the chlorine chemical network, we again used the Meudon code \citep{LePetitetal2006}. The foreground PDR was determined by \citet{LopezSepulcreetal2013a} to have a density of $n~=~10^{2}$~cm$^{-3}$, an FUV irradiation of $\sim1500~G_{0}$ and a cosmic ray ionization rate of $3\times 10^{-16}$~s$^{-1}$. The resulting abundance profiles of Cl, Cl$^{+}$, HCl and \htwoclplus\ are shown in the right-hand panel of Figure~\ref{fig:meudon}. The PDR chemistry model predicts $N$(\htwoclplus)$~=~1.1\times 10^{13}$~cm$^{-2}$, an order of magnitude below the observed value.

An order of magnitude excess of the observed \htwoclplus\ column density over the chemical model predictions, such as we find for \source, has been noted since the recent first identification of the species towards NGC~6334I and Sgr~B2(S) by \citet{Lisetal2010}, who suggested viewing geometry as a possible explanation -- PDR model column densities are commonly given for a face-on viewing angle. The discrepancy was discussed in detail for several sources by \citet{Neufeldetal2012}, but no satisfactory explanation surfaced, although suggestions for future work were given by the authors in their Section~6. Thus, while our data is consistent with no depletion of chlorine from the gas in the tenuous foreground layer, evidently our understanding either of chlorine in diffuse PDRs or of the geometry and conditions in these regions, is not yet complete.

\section{Discussion}\label{sec:discussion}

\subsection{Depletion of chlorine}\label{sec:rayflux}

We find a ratio of elemental volatile chlorine to gas-phase HCl, $X$(Cl)$_{\rm tot}/X$(HCl)$_{\rm gas}$, of $\sim1000$ in \source. Previous studies of the HCl abundance in molecular gas have found ratios in the range $50$ to $640$, typically $\sim400$ \citep[e.g.][]{Schilkeetal1995, Zmuidzinasetal1995, Salezetal1996, NeufeldGreen1994, Pengetal2010}. A depletion factor of $\sim100$ was found for the outflow-shocked region L1157-B1 \citep{Codellaetal2012}.

Our modelling suggests that hydrogen chloride ice is the main chlorine reservoir in protostellar core conditions, containing $90$ to $98$\% of the elemental volatile chlorine. Gas-phase atomic Cl contains most of the remaining $10$\% of the chlorine. 

The above result is independent of our choice of of cosmic ray ionization rate, $\zeta$, although inside of $\sim5000$~AU, the gas-phase abundance of HCl (a minor, but key part in the chlorine budget) can vary substantially depending on this parameter. For $\zeta=10^{-16}$~s$^{-1}$, the gaseous HCl abundance stays similar to that of the outer envelope ($\sim10^{-10}$), however for $\zeta=10^{-14}$~s$^{-1}$ it is $10^{-8}$ up to $10^{5}$~years. This is, at face value, not consistent with our observations, as it would cause extremely strong HCl emission which is not observed. On the other hand, evolving the static $\zeta=10^{-14}$~s$^{-1}$ model beyond $10^{5}$~years leads to a decreasing abundance of HCl in the gas, and it eventually falls below $10^{-10}$ in the inner envelope.

Another possibility is that infall influences the abundances in the inner envelope, keeping more chlorine in HCl ice than is seen in our static models. This may require quite rapid infall (within a few thousand years on $5000$~AU scales) in order to prevent the buildup of a high gas-phase abundance. The poorly known physical structure on $\lesssim2000$~AU scales also impacts our modelling of the chemistry as well as the excitation of HCl, leading to further uncertainty about the HCl abundance on small scales. Another possibility is that HCl is liberated from grain mantles more slowly than the bulk ice. This might also explain the low gas-phase HCl abundance found in the L1157-B1 shock by \citep{Codellaetal2012}, and would point to some -- as yet unknown -- relatively refractory reservoirs of chlorine on grains.

The dominance of HCl ice as a reservoir of volatile Cl warrants a discussion of the relevant grain surface processes. Much chlorine arrives on ice mantles in atomic form, rather than already in HCl, and then subsequently reacts with H to form HCl. Chlorine may also react with \htwo\ also present on an interstellar grain surface, as the barrier for reaction with \htwo\ is measured to be only $2300$~K in the gas phase. On a grain surface, \htwo\ can tunnel through a barrier of up to $4700$~K \citep{TielensHagen1982}. In this way, Cl will act similarly to OH, which also has a low barrier for reaction with \htwo\, and theory and experiments have shown that that reaction is key to interstellar \htwoo\ formation \citep{TielensHagen1982, Obaetal2012}.

Analogous to water solutions of hydrochloric acid, and given the low elemental abundance of chlorine, adsorbed HCl can solvate as a trace ion pair \citep[Cl$^{-}$ and H$_{3}$O$^{+}$, e.g.][]{Hornetal1992}. Theoretical studies suggest this process is energetically allowed on an ice surface and proceeds rapidly by tunnelling at $190$~K \citep{RobertsonClary1995}. Experimental studies show that HCl adsorbs dissociatively at sub-monolayer coverages onto the surface of dense amorphous solid water at temperatures as low as $20$~K \citep{Ayotteetal2011}. As dangling OH bonds are involved -- which will be omnipresent on growing interstellar ice surfaces -- and in view of the long interstellar timescales, we consider solvation likely on a $10$~K icy interstellar grain. Observationally, it is well established that ion-solvation is a key aspect of interstellar ices \citep{Demyketal1998} and experiments have shown that ion-solvation can occur at low temperatures and is promoted by the presence of strong bases such as NH$_{3}$, leading to trapped Cl$^{-}$--NH$_{4}^{+}$ ion pairs \citep{Grimetal1989}. Dipole alignment in ice mantles can further assist in ion-solvation \citep{Balogetal2011}.

Upon warmup, HCl will evaporate close to the \htwoo\ evaporation temperature. This likely involves the relaxation of the water ice matrix, followed by the recombination and evaporation of HCl \citep{Olanrewajuetal2011}. The observed depth of depletion outside of the hot core in \source\ is consistent with such a codesorption scenario, although with the present data we can only place an upper limit of $10^{-8}$ on the gas-phase HCl abundance in the $\sim500$~AU size hot core (or $10^{-7}$ within $100$~AU).

\subsection{Uncertainty in the source luminosity}\label{sec:luminosity}

The $1000$~\lsol\ luminosity of the FIR~4 source model from C09 exceeds the protostellar luminosity of $100$~\lsol\ found by \citet[][F14]{Furlanetal2014}. The latter authors, as well as \citet{LopezSepulcreetal2013b}, attribute this to the lower spatial resolution data and larger photometric annuli used by C09. We assess here the potential impact on our results.

While the far-infrared fluxes of C09 were likely contaminated by the nearby source, FIR~3, the (sub-)mm continuum maps were spatially resolved on a scale comparable to that studied by F14 and must be reproduced by any model. For a luminosity and thus cooling rate decrease of $1$~dex, the corresponding dust temperature decrease is a factor of $1.5$ or $\sim30$\%. The fixed mm flux demands a matching increase in the density of the emission region. Recalculating the HCl~$1-0$ excitation for these conditions implies a change in our derived HCl abundance by at most a factor of two.

The impact of the changed source luminosity on the chemistry is also expected to be small. Lowered temperatures would most of all assist in keeping chlorine locked in ices.

\subsection{The broad component}\label{sec:broad}

The broad component has a combination of relatively large line width ($\sim10$~km/s) and large spatial extent (thousands of AU), while embedded in, or at least projected over, a dense protostellar core. As we discuss below, it is not obvious what the nature of this component is. Kinematical evidence relates the broad HCl component to the CS molecule, mapping of which in turn suggests that this component is extended on a scale of $\gtrsim 2100$~AU. Our constant HCl abundance models suggest a best-fit spherically symmetric radial extent of $7000$~AU. At first glance, such a large line width and large spatial scale make hypotheses other than an outflow seem unlikely.

There is indeed mounting evidence from kinematical and excitation considerations that \source\ indeed hosts a compact outflow \citep{Kamaetal2013, Furlanetal2014, Kamaetalinprep}. Spatially and spectrally resolved data, to be presented in a companion paper, suggest that the outflow axis runs roughly North to South, with lobe sizes of at a few thousand AU. However, the C$^{34}$S map in Figure~\ref{fig:csmap} shows a significant East-West elongation, which seems difficult to explain with such an outflow, unless outflow-driven gas is spilling over the protostellar core surface where the outflow cone breaks out. The C$^{34}$S velocity map shows a slow rotation around the North-South axis, with a typical velocity an order of magnitude below the linewidth.

It has been proposed that a larger outflow from the nearby Class~I source, OMC-2~FIR~3, impacts and shocks the FIR~4 core \citep{Shimajirietal2008}. It seems unlikely, however, that the FIR~3 outflow is responsible for the broad line emission in FIR~4, because the broad C$^{34}$S emission peaks on-source, and because the high-velocity wings of the CO and \htwoo\ lines in FIR~4 are perfectly symmetric around the local $v_{\rm lsr}$. Spatially resolved studies of the high velocity wings of CO are needed to clarify the issue. Previous interferometric observations have had insufficient sensitivity to probe the outflow gas at several tens of km$\cdot$s$^{-1}$.

\subsection{The chlorine isotopic ratio}

Studies of this ratio throughout the Galaxy have typically been consistent with the Solar System value of $3.1$ \citep{Lodders2003}, within large error bars \citep[e.g.][]{Salezetal1996, Pengetal2010, Cernicharoetal2010a}.

Because of the very small Cl isotope mass difference ($6$\%), minimal chemical fractionation is expected, and we determined the isotope ratio via the HCl and \htwoclplus\ isotopolog ratios. For HCl, we find a line flux ratio of $3.2\pm0.1$, which is a robust isotopolog ratio indicator, given the low optical depth of the lines suggested by the hyperfine component ratios of the narrow HCl emission \citep[see also][]{Cernicharoetal2010a}. For \htwoclplus, we found a ratio of $4.3\pm0.8$. The results, summarized in Table~\ref{tab:clratio}, are consistent with the Solar System value and with values measured elsewhere in the Orion star forming region.

\begin{table}[!ht]
\begin{center}
\caption{\clratio\ determinations towards the Orion star forming region. All uncertainties are $1\sigma$.}
\label{tab:clratio}
\begin{tabular}{ l l l }
Source/Region	& 	\clratio & Notes	\\
\hline
\hline
OMC-2~FIR~4 (HCl)			& \ourhclratio		& this work				\\
OMC-2 foreground (\htwoclplus)	& \ourhtwoclplusratio	& this work			\\
OMC-1 position 1		& $2.3^{+1.8}_{-0.8}$	& \citet{Pengetal2010}		\\
OMC-1 position 2		& $2.5^{+0.9}_{-0.7}$	& \citet{Pengetal2010}		\\
OMC-1				& $6.5^{+2.2}_{-2.2}$	& \citet{Salezetal1996}		\\
Orion Bar				& $2.1^{+0.5}_{-0.5}$	& \citet{Pengetal2010}		\\
\hline
Solar System			& $3.1$				& \citet{Lodders2003}		\\
\end{tabular}
\end{center}
\end{table}

\subsection{Impact of the new HCl-\htwo\ excitation rates}\label{sec:rateimpact}

That the difference between the HCl-\htwo\ excitation rate coefficients and the scaled HCl-He ones should impact abundance determinations was noted already by \citet{lanza_near-resonant_2014}. As shown in Figure~\ref{fig:collrates}, the new HCl-\htwo\ hyperfine-resolved collisional excitation rates are roughly a factor of five to ten larger than the previously used, mass-scaled HCl-He rates from \citet{NeufeldGreen1994} and \citet{lanza_collisional_2012}. This suggests that previous estimates of the gas-phase HCl abundance in molecular gas must be re-evaluated to be up to an order of magnitude lower, and correspondingly the typical fraction of elemental chlorine in gas-phase HCl must be around a factor of $10^{-3}$ (a depletion factor of $\sim1000$). Modelling results in Section~\ref{sec:nautilus} show that the strong depletion can be well understood in a framework where elemental chlorine is sequestered into HCl ice, where it remains at least as strongly bound as \htwoo\ itself.

Based on the new excitation rates, the critical density of the HCl~$1-0$ transition is $\sim10^{7}$~cm$^{-3}$. This is accurate within a factor of a few in the temperature range of the new rate coefficients (up to $300$~K).

\section{Conclusions}\label{sec:conclusions}

We carried out a study of chlorine towards the OMC-2~FIR~4 protostellar core, using \emph{Herschel} and CSO observations of HCl and \htwoclplus. Our main findings are listed below.

\begin{enumerate}
\item{We detect the HCl~$1-0$ and $2-1$ transitions in emission with \herschel\ and CSO, and \htwoclplus\ in absorption with \herschel.}
\item{The narrow HCl component (FWHM~$=2.0$~km/s) traces the outer envelope, and the broad one (FWHM~$=10.5$~km/s) a compact central region, possibly outflow-driven gas.}
\item{The HCl data are well modelled with a constant abundance of $X$(HCl)$_{gas}=9\times 10^{-11}$ in \source, corresponding to $\sim 10^{-3}$ of the ISM abundance of elemental chlorine.}
\item{Chemical models show that HCl ice contains $\sim90$ to $98$\% of all volatile chlorine in the source. The second largest reservoir is gas-phase atomic Cl, up to $10$\% of the total. All other species have much lower abundances. In the inner $100$~AU, HCl gas may hold up to $100$\% of volatile chlorine.}
\item{The external irradiation of the \source\ envelope is $G_{0}=415$~ISRF. Elemental chlorine is undepleted in the outermost $300$~AU of the resulting dense PDR. Including this PDR in the source model gives a best-fit $X$(HCl)$_{gas}=7\times 10^{-11}$ in the rest of the source.}
\item{\htwoclplus\ traces a recently discovered diffuse, blueshifted foreground PDR. The observed \htwoclplus\ column density is $1.3\times 10^{14}$~cm$^{-2}$, an order of magnitude above the model prediction of $1.1\times 10^{13}$~cm$^{-2}$.}
\item{Our best estimate of the \clratio\ isotope ratio in OMC-2~FIR~4 is $3.2\pm0.1$ ($1~\sigma$), consistent with other measurements in the Solar System and in the Orion region.}
\item{Newly calculated HCl-\htwo\ hyperfine-resolved collisional excitation rate coefficients exceed previous HCl-He scaled values by up to an order of magnitude at protostellar core temperatures, suggesting that previous estimates of chlorine depletion from the gas should be revisited.}
\end{enumerate}

\begin{acknowledgements}
We would like to thank the anonymous referee for constructive comments that helped to improve the manuscript. We also thank Catherine Walsh, Alexandre Faure, Yulia Kalugina, Laurent Wiesenfeld and Ewine van Dishoeck for useful discussions; Charlotte Vastel for help with molecular data; and Evelyne Roueff for support with the Meudon code. Astrochemistry in Leiden is supported by the Netherlands Research School for Astronomy (NOVA), by a Royal Netherlands Academy of Arts and Sciences (KNAW) professor prize, and by the European Union A-ERC grant 291141 CHEMPLAN. V.W. acknowledges funding by the ERC Starting Grant 3DICE (grant agreement 336474). F.L. and M.L. acknowledge support by the Agence Nationale de la Recherche (ANR-HYDRIDES), contract ANR-12-BS05-0011-01, by the CNRS national program  ``Physique et Chimie du Milieu Interstellaire'' and by the CPER Haute-Normandie/CNRT/Energie, Electronique, Mat\'{e}riaux. Support for this work was provided by NASA (\emph{Herschel} OT funding) through an award issued by JPL/Caltech. We gratefully acknowledge G\"{o}ran Pilbratt for granting \emph{Herschel} Director's Discretionary Time that greatly improved the HIFI data sensitivity. HIFI has been designed and built by a consortium of institutes and university departments from across Europe, Canada and the United States under the leadership of SRON Netherlands Institute for Space Research, Groningen, The Netherlands and with major contributions from Germany, France and the US. Consortium members are: Canada: CSA, U.Waterloo; France: CESR, LAB, LERMA, IRAM; Germany: KOSMA, MPIfR, MPS; Ireland, NUI Maynooth; Italy: ASI, IFSI-INAF, Osservatorio Astrofisico di Arcetri-INAF; Netherlands: SRON, TUD; Poland: CAMK, CBK; Spain: Observatorio Astronómico Nacional (IGN), Centro de Astrobiología (CSIC-INTA). Sweden: Chalmers University of Technology - MC2, RSS \& GARD; Onsala Space Observatory; Swedish National Space Board, Stockholm University - Stockholm Observatory; Switzerland: ETH Zurich, FHNW; USA: Caltech, JPL, NHSC. The Caltech Submillimeter Observatory is operated by the California Institute of Technology under cooperative agreement with the National Science Foundation (AST-0838261). Based on analysis carried out with the CASSIS software. CASSIS has been developed by IRAP-UPS/CNRS.
\end{acknowledgements}

\bibliographystyle{aa}
\bibliography{omc2chlorine}

\begin{appendix}

\section{Hyperfine excitation of HCl by H$_2$}\label{apx:collrates}

Rate coefficients for rotational excitation of HCl($^1\Sigma^+$) by collisions with \htwo\ molecules have been computed by \citet{lanza_taux} for temperatures ranging from 5 to 300~K. The rate coefficients were derived from extensive quantum calculations using a new accurate potential energy surface obtained from highly correlated \emph{ab initio} approaches \citep{lanza_near-resonant_2014}. 

However, in these calculations, the hyperfine structure of HCl was neglected. To model the spectrally resolved HCl emission from molecular clouds, hyperfine resolved rate coefficients are needed. In this appendix, we present the calculations of HCl--\htwo\ hyperfine resolved rate coefficients from the rotational rate coefficients of \citet{lanza_taux}. Note that, for rotational levels, we use here the lowercase $j$ instead of the astronomical $J$ notation used in the main body of the paper.

\subsection{Methods}

In HCl, the coupling between the nuclear spin ($I_1=3/2$) of the chlorine atom and the molecular rotation results in a weak splitting of each rotational level $j_1$ into 4 hyperfine levels (except for the $j_1=0$ level which is split into only 1 level and for the $j_1=1$ level which is split into only 3 levels). Each hyperfine level is designated by a quantum number $F_1$ ($F_1=I_1+j_1$) varying between $|I_1-j_1|$ and $I_1+j_1$. In the following, $j_2$ designates the rotational momentum of the H$_2$ molecule.

In order to get HCl--\htwo\ hyperfine resolved rate coefficients, we extend the Infinite Order Sudden (IOS) approach for diatom-atom collisions \citep{Faure:12} to the case of diatom-diatom collisions.

Within the IOS approximation, inelastic rotational
rate coefficients $k^{IOS}_{j_1,j_2 \to j_1',j_2'}(T)$ can be calculated from the ``fundamental'' rates (those
out of the lowest $j_1=0,j_2=0$ channel) as follows \citep[e.g][]{alexander79}:
\begin{eqnarray} 
k^{IOS}_{j_1,j_2 \to j'_1,j'_2}(T)   & = &
 \sum_{L_1,L_2}(2j'_2+1)(2j'_1+1)  \left(\begin{array}{ccc}
j_1 & L_1 & j'_1 \\ 
0 & 0 & 0
\end{array}\right)^{2} \nonumber \\
& & \times 
 \left(\begin{array}{ccc}
j_2 & L_2 & j'_2 \\ 
0 & 0 & 0
\end{array}\right)^{2} 
 k^{IOS}_{0,0 \to L_1,L_2}(T)
\label{iosrot}
\end{eqnarray}

Similarly, IOS rate coefficients amongst hyperfine structure levels can be obtained from the
$k^{IOS}_{0,0 \to L_1,L_2}(T)
$ rate coefficients using the following formula:
\begin{eqnarray} \label{REEQ}
k^{IOS}_{j_1,F_1,j_2 \to j'_1,F'_1,j'_2}(T)   & = & 
\sum_{L_1,L_2}(2j'_2+1)(2j_1+1)(2j'_1+1)  \nonumber \\
& & \times (2F'_1+1) 
 \left(\begin{array}{ccc}
j_2 & L_2 & j'_2 \\ 
0 & 0 & 0
\end{array}\right)^{2}  
 \left(\begin{array}{ccc}
j'_1 & j_1 & L_1 \\ 
0 & 0 & 0
\end{array}\right)^{2} \nonumber \\
& & \times \left\{\begin{array}{ccc}
L_1 & F_1 & F'_1 \\ 
I_1 & j'_1 & j_1 
\end{array}\right\}^{2}
 k^{IOS}_{0,0 \to L_1,L_2}(T)
\end{eqnarray}
where $\left( \quad \right)$ and $\left\{ \quad \right\}$ are
respectively the ``3-j'' and ``6-j'' Wigner symbols.\\

The IOS approximation is expected to be moderately accurate at low temperature. As suggested by \citet{NeufeldGreen1994}, we could improve the accuracy by computing the hyperfine rate coefficients as:

\begin{equation}
\label{scaling}
k^{SIOS}_{j_1,F_1,j_2 \to j'_1F'_1,j'_2}(T)=\frac{k^{IOS}_{j_1,F_1,j_2 \to j'_1,F'_1,j'_2}(T)}
{k^{IOS}_{j_1,j_2 \to j'_1,j'_2}(T)} k^{CC}_{j_1,j_2 \to j'_1,j'_2}(T)
\end{equation}
using the CC rate coefficients $k^{CC}_{0,0 \to L_1,L_2}(T)$ of \cite{lanza_taux} for the IOS
``fundamental'' rates in
Eqs.~\ref{iosrot}-\ref{REEQ}. $k^{CC}_{j_1,j_2 \to j'_1,j'_2}(T)$ are the rotational rate coefficients also taken from \citet{lanza_taux}. We named the method `SIOS" for scaled IOS.

In addition, fundamental excitation rates
$k^{CC}_{0,0 \to L_1,L_2}(T)$ were replaced by the de-excitation
fundamental rates using the detailed balance relation:
\begin{equation}
k^{CC}_{0,0 \to L_1,L_2}(T) = (2L_1+1)(2L_2+1) k^{IOS}_{L_1,L_2 \to 0,0}(T)
\end{equation}
This procedure is found to significantly improve the results
at low temperature due to important threshold effects. 

Hence, we have determined hyperfine HCl--H$_2$ rate coefficients using the computational scheme described above for temperature ranging from 5 to 300K. We considered transitions between the 28 first hyperfine levels of HCl ($j$, $j' \le 7$) due to collisions with para-H$_2(j_2=0)$ and ortho-H$_2(j_2=1)$. The present approach has been shown to be accurate, even at low temperature, and has also been shown to induce almost no inaccuracies in radiative transfer modeling compared to more exact calculations of the rate coefficients \citep{Faure:12}.
 
\subsection{Results}

\begin{center}
\begin{figure}
\includegraphics[width=8cm]{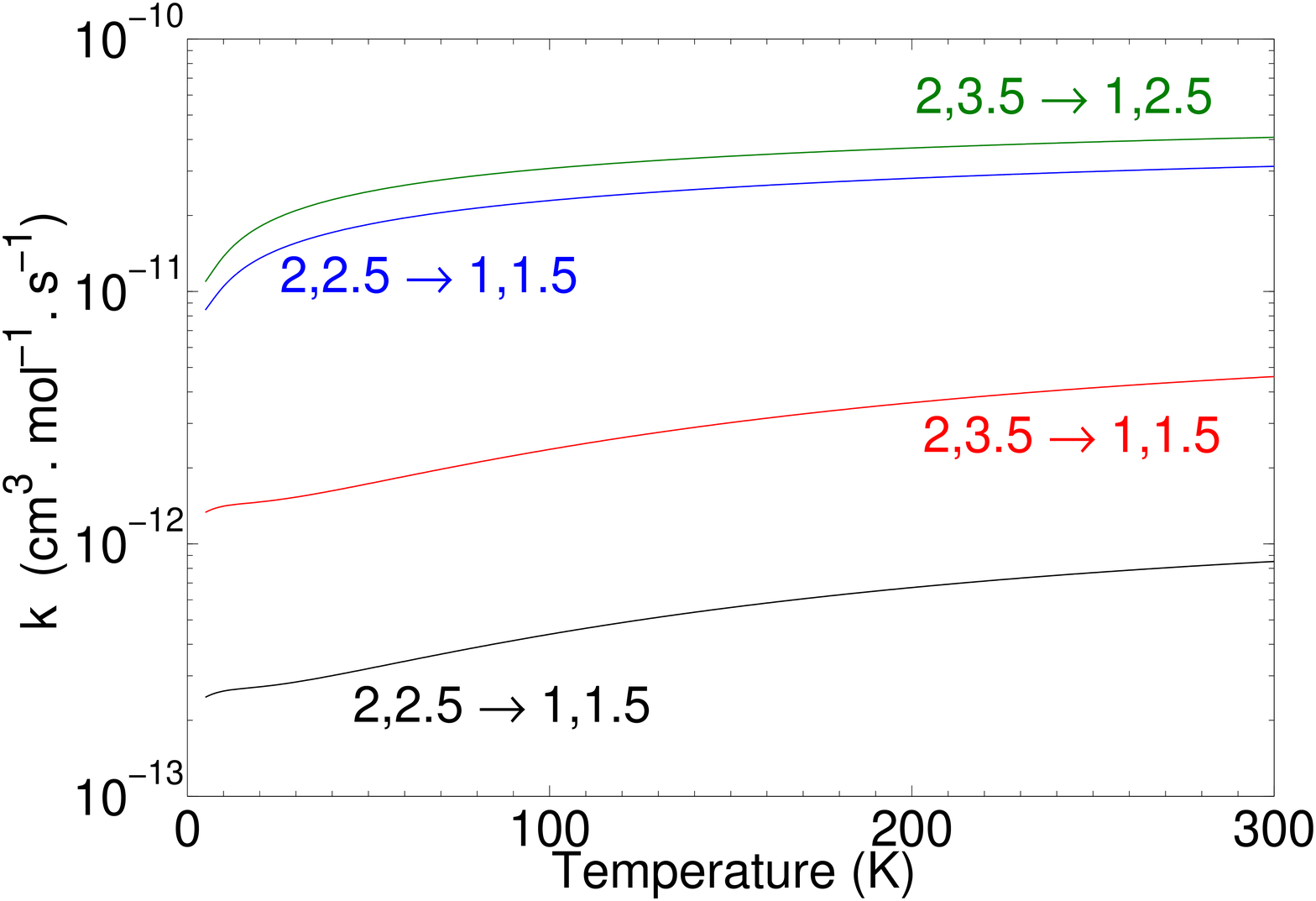}
\includegraphics[width=8cm]{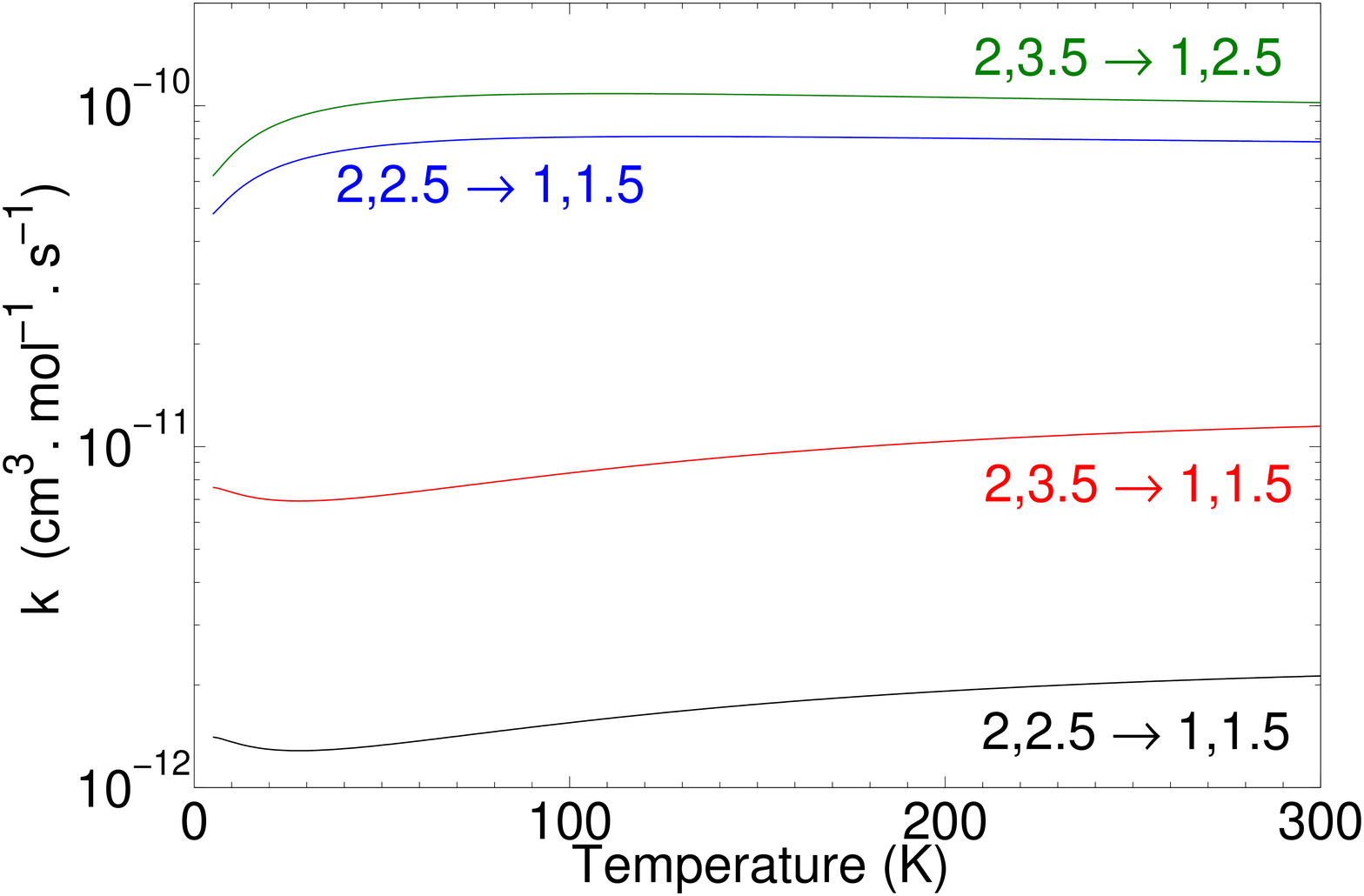}
\caption{The temperature dependence of the hyperfine resolved HCl--\phtwo\ (upper panel) and HCl--\ohtwo\ (lower panel) rate coefficients for HCl($j_{1}=2,F \to j_{1}'=1,F'$) transitions.}
\label{taux}
\end{figure}
\end{center}

The complete set of (de)excitation rate coefficients is available on-line from the LAMDA\footnote{http://www.strw.leidenuniv.nl/~moldata/} \citep{schoier:05} and BASECOL\footnote{http://basecol.obspm.fr/} \citep{Dubernet:13} websites. For illustration, Fig. \ref{taux} depicts the evolution of para- and ortho-H$_2$ rate coefficients as a function of temperature for HCl($j=2,F \to j'=1,F'$) transitions. 

First of all and as already discussed in \citet{lanza_taux}, para- and ortho-H$_2$ rate coefficients differ significantly, the rate coefficients being larger for ortho-H$_2$ collisions.
One can also clearly see that there is a strong propensity in favour of $\Delta j_1= \Delta F_1$ transitions for both collisions with para- and ortho-H$_2$. This trend is the usual trend for such a molecule \citep{Roueff:13}.

\begin{table*}
\caption{A comparison between present hyperfine rate coefficients for ortho- and para-\htwo, and those of LL12 for He. The rates are in units of cm$^3$ mol$^{-1}$ s$^{-1}$ \label{tab1}}
\begin{tabular}{c|ccc|ccc|ccc}
      & \multicolumn{3}{c|}{10 K} & \multicolumn{3}{c|}{100 K} & \multicolumn{3}{c}{300 K} \\
      \hline
$j_{1}$, $F_{1}$ $\rightarrow$ $j_{1}$', $F_{1}$' & p-H$_2$ & o-H$_2$ & He$\times$1.38 & p-H$_2$ & o-H$_2$ & He$\times$1.38 & p-H$_2$ & o-H$_2$ & He$\times$1.38 \\
      \hline
      \hline
1, 1.5 $\rightarrow$ 0, 1.5 & 5.54e-11 & 1.53e-10 & 3.97e-12 & 6.99e-11 & 1.53e-10 & 1.99e-11 & 5.19e-11 & 1.13e-10 & 3.45e-11 \\
2, 2.5 $\rightarrow$ 0, 1.5 & 2.01e-11 & 3.33e-11 & 2.64e-12 & 2.31e-11 & 5.07e-11 & 8.82e-12 & 2.53e-11 & 6.15e-11 & 2.40e-11 \\
2, 3.5 $\rightarrow$ 1, 1.5 & 1.41e-12 & 7.35e-12 & 3.38e-12 & 2.37e-12 & 8.37e-12 & 3.35e-12 & 4.60e-12 & 1.15e-11 & 3.55e-12 \\
2, 3.5 $\rightarrow$ 1, 2.5 & 1.38e-11 & 7.17e-11 & 1.54e-11 & 3.07e-11 & 1.08e-10 & 2.65e-11 & 4.08e-11 & 1.02e-10 & 4.27e-11 \\
3, 3.5 $\rightarrow$ 0, 1.5 & 9.31e-12 & 1.08e-11 & 2.28e-12 & 8.72e-12 & 1.05e-11 & 3.06e-12 & 9.65e-12 & 1.16e-11 & 3.59e-12 \\
3, 3.5 $\rightarrow$ 1, 1.5 & 6.28e-12 & 1.58e-11 & 2.04e-12 & 8.62e-12 & 2.47e-11 & 6.66e-12 & 1.39e-11 & 3.73e-11 & 1.92e-11 \\
3, 2.5 $\rightarrow$ 1, 1.5 & 5.90e-12 & 1.48e-11 & 1.59e-12 & 8.07e-12 & 2.31e-11 & 5.56e-12 & 1.28e-11 & 3.42e-11 & 1.69e-11 \\
3, 2.5 $\rightarrow$ 2, 3.5 & 3.85e-13 & 2.27e-12 & 1.45e-12 & 7.31e-13 & 3.13e-12 & 2.32e-12 & 2.37e-12 & 6.32e-12 & 6.15e-12 \\
3, 4.5 $\rightarrow$ 2, 2.5 & 4.56e-13 & 2.69e-12 & 1.84e-12 & 8.05e-13 & 3.44e-12 & 2.29e-12 & 2.51e-12 & 6.65e-12 & 3.99e-12 \\
\hline
\end{tabular}
\end{table*}

Finally, we compare in Table~\ref{tab1} our new  hyperfine HCl--\htwo\ rate coefficients with the HCl--He ones calculated by \citet{lanza_collisional_2012} which are scaled by a factor $1.38$ to account for the mass difference (see Figure~\ref{fig:collrates} for a visual comparison). Indeed, collisions with helium are often used to model collisions with \emph{para}-\htwo. It is generally assumed that rate coefficients with \emph{para}-\htwo($j_{2}=0$) should be larger than He rate coefficients owing to the smaller collisional reduced mass.   

As one can see, the scaling factor is clearly different from $1.38$. The ratio varies with the transition considered and also with the temperature for a given transition. The ratio may be larger than a factor $10$. This comparison indicates that accurate rate coefficients with para-H$_2$($j_{2}=0$) and ortho-H$_2$($j_{2}=1$) could not be obtained from He rate coefficients. HCl molecular emission analysis performed with HCl--He rate coefficients result in large inaccuracies in the HCl abundance determination.

\section{CASSIS fitting of \htwoclplus}\label{apx:h2clplus}

\begin{figure*}[!ht]
\includegraphics[clip=,width=1.0\linewidth]{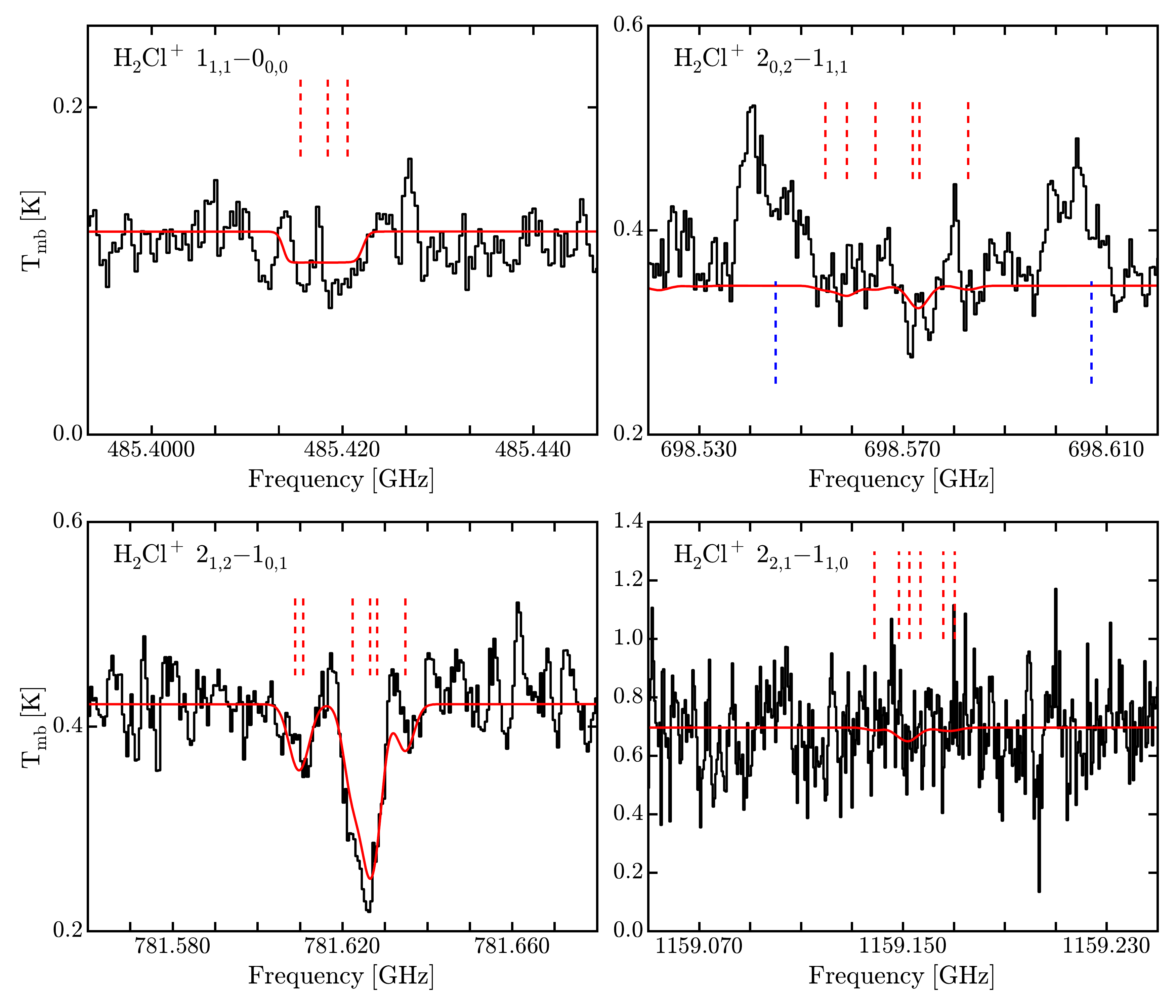}
\caption{The four \htwoclplus\ lines (black) used in the LTE model fitting with CASSIS, and the best-fit model (red). All spectra are corrected for the foreground PDR velocity of $9.4$~km/s. The dashed red lines show the hyperfine components of \htwoclplus\ transitions, while dashed blue lines in the top right panel indicate the native frequencies of C$_{2}$H transitions. The other \htwoclplus\ transitions do not have any lines nearby that were listed as detections in the spectral survey of \citet{Kamaetal2013}.}
\label{fig:h2clplus_cassis}
\end{figure*}

In Figure~\ref{fig:h2clplus_cassis}, we show the \htwoclplus\ lines used in the LTE fitting with the CASSIS software and the best-fit model resulting from the $\chi^{2}$ minimization. The hyperfine components and other lines detected nearby are also shown. The data are all from the HIFI spectral survey of \source\ \citep{Kamaetal2013}.

\end{appendix}

\end{document}